\documentclass[11pt]{article}
\newcommand{\beq}{\begin{equation}}
\newcommand{\eeq}{\end{equation}}
\newcommand{\beqa}{\begin{eqnarray}}
\newcommand{\eeqa}{\end{eqnarray}}
\newcommand{\f}{\frac}

\newcommand{\sn}{{\rm sn}}
\newcommand{\cn}{{\rm cn}}
\newcommand{\dn}{{\rm dn}}
\newcommand{\ds}{{\rm ds}}
\newcommand{\cs}{{\rm cs}}
\newcommand{\ns}{{\rm ns}}

\newcommand{\nc}{{\rm nc}}
\newcommand{\dc}{{\rm dc}}
\newcommand{\am}{{\rm am}}
\newcommand{\Z}{{\rm Z}}
\newcommand{\m}{{\rm m}\,}

\newcommand{\D}{\Delta}
\newcommand{\sech}{{\rm sech}}
\newcommand{\sss}{{\vspace{.2in}}}

\begin{document}
~\hfill{\footnotesize SUNYB/03-04, IOP-BBSR/03-12}
\vspace{.5in}
\begin{center}
{\LARGE {\bf Local Identities Involving Jacobi Elliptic Functions}}
\end{center}
\vspace{.3in}
\begin{center}
{\Large{\bf  \mbox{Avinash Khare}
 }}\\
\noindent
{\large Institute of Physics, Sachivalaya Marg, Bhubaneswar 751005, India}
\end{center}
\begin{center}
{\Large{\bf  \mbox{Arul Lakshminarayan}
 }}\\
\noindent
{\large Physical Research Laboratory, Navrangpura, Ahmedabad 380009, India}
\end{center}
\begin{center}
{\Large{\bf  \mbox{Uday Sukhatme}
 }}\\
\noindent
{\large Department of Physics, State University of New York at Buffalo, Buffalo, NY 14260, U.S.A. }\\
\end{center}
\vspace{.9in}
{\bf {Abstract:}} We derive a number of local identities of arbitrary
rank involving Jacobi elliptic functions and use them to obtain
several new results. First, we present an alternative, simpler
derivation of the cyclic identities discovered by us recently, along
with an extension to several new cyclic identities of arbitrary
rank. Second, we obtain a generalization to cyclic identities in which
successive terms have a multiplicative phase factor $\exp (2i\pi/s)$,
where $s$ is any integer. Third, we systematize the local identities
by deriving four local ``master identities'' analogous to the master
identities for the cyclic sums discussed by us previously. Fourth, we
point out that many of the local identities can be thought of as exact
discretizations of standard nonlinear differential equations satisfied
by the Jacobian elliptic functions. Finally, we obtain explicit
answers for a number of definite integrals and simpler forms for
several indefinite integrals involving Jacobi elliptic functions.

\newpage
\section{Introduction}

In a recent paper \cite{ksjmp}, (henceforth referred to as I), we have
given many new mathematical identities involving the Jacobi elliptic functions
$\sn \,(x,\m)$, $\cn \,(x,\m)$, $\dn \,(x,\m)$, where $\m$ is the elliptic
modulus parameter $( 0\leq \m\leq 1)$. The
functions $\sn \,(x,\m)$, $\cn \,(x,\m)$, $\dn \,(x,\m)$ are doubly periodic
functions with periods $(4K(\m), i2K'(\m))$, $(4K(\m), i4K'(\m))$,
$(2K(\m), i4K'(\m))$,
respectively \cite{abr,gr}.
Here, $K(\m) \equiv \int_0^{\pi/2} d\theta [1-\m\sin^2 \theta]^{-1/2}$ denotes the complete elliptic integral of the first kind, and
$K'(\m)\equiv K(1-\m)$. The $\m =0$ limit gives
$K(0)=\pi /2$ and trigonometric functions: $\sn(x,0)=\sin x, ~\cn(x,0)=\cos x,
~\dn(x,0)=1$. The $\m \rightarrow 1$ limit gives $K(1) \rightarrow \infty$ and
hyperbolic functions: $\sn(x,1) \rightarrow \tanh x,
~\cn(x,1) \rightarrow \sech \,x, ~\dn(x,1) \rightarrow \sech\, x$. For simplicity, from now on
we will not explicitly display the modulus parameter $\m$ as an argument of the
Jacobi elliptic functions.

The identities discussed in ref. I are all cyclic with the
arguments of the Jacobi functions in successive terms
separated by either $2K(\m)/p$ or $4K(\m)/p$, where $p$ is an integer. 
Each $p$-point
identity of rank $R$ involves on its left hand side
a cyclic homogeneous polynomial in Jacobi elliptic functions of degree
$R$ with $p$ equally spaced arguments. The separation is $2K(\m)/p$ or $4K(\m)/p$ depending on whether the period of any term on the left hand side is $2K(\m)$ or $4K(\m)$. 
In another recent publication \cite{kls2} (referred to as II),
we presented rigorous
mathematical proofs valid for arbitrary $p$ and $R$ even though, for simplicity,
we only presented identities of low rank.
In ref. II, we classified the identities
into four types, each with its own ``master identity'' which we
proved using a combination of the Poisson summation formula and the
special properties of elliptic functions \cite{kls2,kls1}.
We also provided a rigorous derivation
of cyclic identities with successive terms having alternating signs.

In this paper, we provide several generalizations of the results discussed in refs. I and II. Here, our approach is different and involves the use of ``local" identities which focus on just any one term in a cyclic identity. This term involves a product of Jacobi elliptic functions and is expressed via the local identity as the sum of many terms of lower rank. The purpose of this paper is to derive and make use of a number of local identities for Jacobi
elliptic functions. These local identities form the building blocks for cyclic as well as much more general identities. For instance, adding $p$ local identities with equally spaced arguments permits us to re-derive cyclic identities. More generally, taking $p$ local identities with a phase $(-1)^{(j-1)}=e^{(j-1)i\pi}$ and summing over the index $j$ gives previously derived identities in which successive terms have alternative signs. Finally, as discussed below, the generalization to taking $p$ local identities with an even more general phase $\exp (2i(j-1)\pi/s)$, where $s$ is any integer and summing over the index $j$ yields interesting new identities in which successive terms have different weights.  Also, while in principle we were
able to prove the general form for identities of arbitrary rank in refs. I and II, in practice it was very difficult
to obtain the explicit coefficients in these identities. The use of local identities permits evaluation of these coefficients. As a byproduct, a number of definite integrals involving Jacobi elliptic functions can be explicitly evaluated and a number of indefinite integrals can be expressed in simpler form. Finally, we show that some special linear combinations of various identities mentioned above have a particularly simple right hand side, and we give several illustrative examples. 

To clarify the above ideas about the approach used in this paper, consider as an example, one specific basic local identity derived here:
\beq
\dn^2(y)\dn(y+a)
=-\cs^2(a)\dn(y+a)+\ds(a)\ns(a)\dn(y)-\m\cs(a)\cn(y)\sn(y)~.
\eeq
Choosing $y=x+(j-1)2K(\m)/p$ with $j=1,2,...,p$ actually corresponds to $p$ identities, one for each value of the integer $j$. 
Taking $a=r2K(\m)/p$, where $r$ is an integer which is less than $p$ and coprime to it, and summing over $j$ yields the cyclic identity
\beq
\sum_{j=1}^{p} d_j^2 d_{j+r} = \sum_{j=1}^{p} \left [\frac{A}{2}d_j-\m\cs(a)s_j c_j \right]~,
\eeq
where the coefficient $A$ is given by $A = 2 [\ds(a)\ns(a)-\cs^2(a)]$ and we have used the notation
\beq\label{1a}
d_j \equiv \dn(x+(j-1)2K(\m)/p,m),~s_j \equiv \sn(x+(j-1)2K(\m)/p,m),
~c_j \equiv \cn(x+(j-1)2K(\m)/p,m)~.
\eeq
Similar manipulations using $a=-r2K(\m)/p$ yield cyclic identities for expressions like $\sum_{j=1}^{p} d_j^2 [d_{j+r} \pm d_{j-r}]$. The result with the negative sign is new and will be discussed later in this paper. The result with the positive sign is
\beq\label{1}
\sum_{j=1}^{p} d_j^2 \left [d_{j+r} + d_{j-r} \right ] = A\sum_{j=1}^{p} d_j~,
\eeq
and is one of the cyclic identities derived in ref. II by more complicated techniques.

Further, while it was clear
from refs. I and II that the identities of arbitrary (odd) rank
which are generalizations of eq. (\ref{1}) must have
the structure
\beq\label{2}
\sum_{j=1}^{p} d_j^{2n} \left [d_{j+r}+d_{j-r} \right ]
= A_1 \sum_{j=1}^{p} d^{2n-1}_j~
+...+A_n \sum_{j=1}^{p} d_j~,
\eeq
we were unable to obtain the coefficients $A_1,...,A_n$. Here, we will obtain explicit expressions for the coefficients.
In ref. II, we were able to obtain the analogue of 
identity (\ref{1}) with
alternating signs given by
\beq\label{3}
\sum_{j=1}^{p} (-1)^{j-1} d_j^2 \left [d_{j+r}+d_{j-r} \right ]
= A\sum_{j=1}^{p} (-1)^{j-1} d_j~,~~
A=2 \left [\ds(a)\ns(a)+\cs^2(a) \right ]~,~~a=\f{r2K}{p}~.
\eeq
Here, we will obtain identities with
more general weights like
\beq\label{4}
\sum_{j=1}^{p} \omega^{j-1} d_j^2[d_{j+r}+d_{j-r}]~,~~
\omega =\exp(\f{2i\pi}{s})~,
\eeq
where $\omega$ is the ${s}^{\rm th}$ root of unity, with $s$ being any
integer ($<p$) and $p$ being $0$ mod $s$.  Finally, in ref. II we had
obtained MI-II (class II master identity, also see Sec. 4 below)
identities like
\beq\label{5}
\sum_{j=1}^{p} d_j^2 d^2_{j+r} = -2\cs^2 (a)\sum_{j=1}^{p} d_j^2
+\frac{p}{2K}\left (\int_{0}^{2K} \dn^2 (t) \dn^2 (t+a) dt +4E\cs^2 (a) \right )~,~~
a=\f{r2K}{p}~,
\eeq
where $E$ is the complete elliptic integral of
second kind \cite{gr}. The approach in this paper will permit an evaluation of
the definite integral on the right hand side.

The plan of this paper is as follows. In Sec. 2, we state
several local identities and indicate how they are derived. It may be noted
here that each identity has an integer label $R$ indicating the rank of the
identity, i.e. the left hand side of the identity is a homogeneous
polynomial of degree $R$.
We also show here that linear combinations of cyclic identities often yield simpler results. In Sec. 3, we
use the local identities of rank 2, 3, 4 recursively to obtain local
identities of arbitrary odd and even rank, using which one can
immediately obtain the corresponding cyclic identities with weight factors $\omega$.
In Sec. 4, we provide a unified framework for the local identities
by deriving four master local identities, from which all the
identities can be derived in an alternative manner without using addition formulas. In Sec.
5, we concentrate on those identities of ref. II in which one of the
terms on the right hand side is a definite integral (which we were
previously unable to evaluate). Using our local identities, we show that one
can obtain cyclic identities where all the terms on the right hand
side are now explicitly known. In fact, we show that by starting from any given local identity, the
indefinite integral of the left hand side of this identity can be
analytically obtained in terms of well known integrals of Jacobi
elliptic functions and indefinite elliptic integrals of the first,
second and third kind \cite{gr,bf}. We would like to re-emphasize
that most of these integrals do not seem to be known in the
literature. In Sec. 6, we discuss continuum limits of the local and
cyclic identities, showing that these degenerate to standard
differential equations or integral formulas. Sec. 7 contains
conclusions and a discussion of some open problems. All local identities
of ranks 2, 3 and 4 are presented in Appendices A, B and
C respectively. A few local identities of rank 5 and arbitrary rank are
presented in Appendices D and E respectively. Several simple results
obtained by taking 
suitable linear combinations of cyclic identities are given in Appendix F.
Many new definite and indefinite integrals are given in
Appendices G and H respectively.

\section{The Basic Local Identities}

In this section we shall obtain several basic local identities. These
identities are easily derived using the well-known addition formulas for the $\sn, ~\cn,
 ~\dn$ functions \cite{abr,gr}:
\beq\label{2.1}
\dn(a+b)=\f{\dn(a)\dn(b)-\m\cn(a)\cn(b)\sn(a)\sn(b)}{1-\m\sn^2(a)\sn^2(b)}~,
\eeq
\beq\label{2.2}
\cn(a+b)=\f{\cn(a)\cn(b)-\dn(a)\sn(a)\dn(b)\sn(b)}{1-\m\sn^2(a)\sn^2(b)}~,
\eeq
\beq\label{2.3}
\sn(a+b)=\f{\sn(a)\cn(b)\dn(b)+\cn(a)\dn(a)\sn(b)}{1-\m\sn^2(a)\sn^2(b)}~.
\eeq
We shall also use the addition formula for the Jacobi zeta
function given by
\beq\label{2.4}
\Z(a+b)=\Z(a)+\Z(b)-\m\sn(a)\sn(b)\sn(a+b)~.
\eeq

One of the simplest, local, rank two identities is
\beq\label{2.9}
\dn(x) \dn(x+a) = \dn(a)+\cs(a)[\Z(x+a)-\Z(x) -\Z(a)]~,
\eeq
which is easily proved by algebraic simplification after
using the addition formulas (\ref{2.1}) to (\ref{2.4}).
The power of this local identity can be appreciated by the fact that 
we can immediately derive the cyclic identity
\beq\label{2.6}
\sum_{j=1}^{p} d_j d_{j+1}
= p\left [\dn(\f{2K}{p})-\cs(\f{2K}{p})\Z(\f{2K}{p}) \right ]~,
\eeq
which was obtained in ref. II. This is done by writing local identities like
(\ref{2.9}) with $x$ being replaced by $x+a,~ x+2a,\ldots,~x+(p-1)a$ and
choosing $a=2K/p$. On adding these identities and noting that $\dn(x)$ has
a period $2K$, we then immediately obtain the cyclic identity
(\ref{2.6}). Here $p$ denotes the number of subdivisions of the period at
which Jacobi elliptic functions $\dn(x)$ are evaluated.
A generalization of
this identity to $r^{\rm th}$ neighbours is 
immediate, i.e. on choosing $a=r2K/p$
(where $r$ is coprime to and less than $p$), we obtain
the more general identity
\beq\label{2.7}
\sum_{j=1}^{p} d_j d_{j+r}
= p[\dn(a)-\cs(a)\Z(a)]~,~~a =\f{r2K}{p}~,
\eeq
which was also obtained in ref. II.

We can immediately obtain a local identity for $\dn(x) \dn(x-a)$ by changing $a$ to $-a$ and
recognizing the fact that while $\cn(a),\dn(a)$ are even functions of $a$,
the functions $\sn(a), \Z(a)$ are odd:
\beq\label{2.99}
\dn(x) \dn(x-a) = \dn(a)-\cs(a)[\Z(x-a)-\Z(x) +\Z(a)]~.
\eeq
Adding and subtracting eqs. (\ref{2.9}) and (\ref{2.99}) yields alternative simple expressions:
\beq\label{2.5}
\dn(x)  [\dn(x+a)+\dn(x-a) ]
= 2 \dn(a) +\cs(a)[\Z(x+a)-\Z(x-a) -2\Z(a)]~,
\eeq
and
\beq\label{2.8}
\dn(x)  [\dn(x+a)-\dn(x-a)]
= \cs(a)[\Z(x+a)+\Z(x-a)-2\Z(x)]~.
\eeq

If we now consider the local identities analogous to
(\ref{2.9}) with
$x$ being replaced by $x+a$, $x+2a$,...,$x+(p-1)a$, multiply them in turn 
by $\omega$, $\omega^2$,..., $\omega^{p-2}$, $\omega^{p-1}$
respectively and add to the local identity (\ref{2.9}), then
we obtain the remarkable identity
\beq\label{2.10}
\sum_{j=1}^{p} \omega^{j-1} d_j d_{j+r}
= p [\dn(a)-\cs(a)\Z(a)]\delta_{s1}
- \left( 1-\frac{1}{\omega} \right ) \cs(a)
\sum_{j=1}^{p} \omega^{j-1} \Z_j~,
\eeq
where $a=r2K/p$. The phase $\omega$ is as given by eq. (\ref{4}) with $s < p$ and $p$ being
$0$ mod $s$.
For the special case $s=1$ we recover
the cyclic identity (\ref{2.7}) with all terms on the left hand side having positive signs. For $s=2$, eq. (\ref{2.10}) gives the cyclic identity
with terms having alternating signs as obtained in ref. II. Thus the local identities are very
basic in the sense that once they are known, then the corresponding cyclic
identities with and without arbitrary weight $\omega$ are immediately
obtainable.
It is worth emphasizing here that
the cyclic identities with
arbitrary weight are new.

Proceeding in the same way, we have derived all possible local identities of 
rank two, three and four. They are given in Appendices A, B and C respectively.
Some examples of local identities of rank 5 are given in Appendix D. 
By following the procedure explained above, in each case it is easy to obtain the
corresponding cyclic identities with weights $\omega$.

One advantage of the local identities approach is that for the MI-II type of
cyclic identities, the right hand side is explicitly known. In this
context, it is worth mentioning
that in ref. II (also see \cite{kls1}) we had obtained several MI-II cyclic
identities in which one of the terms on the right hand side is a definite
integral. For example, one of the cyclic MI-II identities obtained in ref. II
is given by eq. (\ref{5}). However, if we take the local identity
(\ref{c15}) given in Appendix C, and use the procedure described above,
we find a simpler, more elegant form for this cyclic identity
\beq\label{2.10a}
\sum_{j=1}^{p} d_j^2 d^2_{j+r} = -2\cs^2(a)\sum_{j=1}^{p} d_j^2
+p[\cs^2(a)+\ds^2(a)-2\cs(a)\ds(a)\ns(a)\Z(a)]~,~~a=\f{r2K}{p}~.
\eeq
Proceeding in this way and using various local identities obtained in this
paper, the corresponding MI-II cyclic identities are easily written where the
constant on the right hand side is now explicitly known and is not just formally
expressed as an unevaluated definite integral.

Let us now consider the following cyclic identity of rank two
\beq\label{2.11}
\m\cn(x)[\sn(x+a)-\sn(x-a)]
=2\ns(a)\dn(x)-\ds(a) [\dn(x+a)+\dn(x-a)]~,
\eeq
which is easily derived by using eqs. (\ref{2.1}) to (\ref{2.4}). 
>From here we easily obtain the following cyclic identity with weighted terms  
\beq\label{2.12}
\sum_{j=1}^{p} \m \omega^{j-1} c_j[s_{j+r}-s_{j-r}]
=2 \left [\ns(a)-\cos(\f{2\pi}{s})\ds(a) \right ] \sum_{j=1}^{p} \omega^{j-1} d_j~,
\eeq
where $a=r2K/p$.
Some examples of new cyclic identities (with weighted terms) of rank 3 and 4 are:
\beqa\label{2.13}
&& \m\sum_{j=1}^{p} \omega^{j-1} d_j
 [c_{j+r}s_{j+r}-c_{j-r}s_{j-r}]
= 2p[\cs(a)-\ds(a)\ns(a) \Z(a)]\delta_{s1} \nonumber \\
&&\hspace{1.in} -2i\sin (\f{2\pi}{s})\ds(a)\ns(a) \sum_{j=1}^{p} \omega^{j-1} \Z_j 
-2 \cos (\f{2\pi}{s}) \cs(a) \sum_{j=1}^{p} \omega^{j-1} d_j^2~,
\eeqa
\beq\label{2.14}
 \sum_{j=1}^{p} \omega^{j-1} d_j
 [c_{j+r}d_{j+r}-c_{j-r}d_{j-r}]
=2 \cs(b) \cos (\f{2\pi}{s}) \sum_{j=1}^{p} \omega^{j-1} s_j d_j
-2i\sin (\f{2\pi}{s})\ds(b)\ns(b) \sum_{j=1}^{p} \omega^{j-1} c_j~,
\eeq
\beqa\label{2.15}
&&\!\!\!\!\!\!\m \sum_{j=1}^{p} \omega^{j-1} c_j
 [c_{j+r}s_{j+r}d_{j+r}-c_{j-r}s_{j-r}d_{j-r}]
=-2i \sin (\f{2\pi}{s}) \cs(b) \ns(b) 
\sum_{j=1}^{p} \omega^{j-1} s_j d_j \nonumber \\
&&\!\!\!\!\!\!-2\m\cos (\f{2\pi}{s})\ds(b)
\sum_{j=1}^{p} \omega^{j-1} c^3_j
+2\ds(b)
\left [(\m+\cs^2(b)) \cos(\f{2\pi}{s})-\cs(b) \ns(b) \right ]
\sum_{j=1}^{p} \omega^{j-1} c_j
,~b =\frac{r4K}{p}.
\eeqa
It may be noted that the identities (\ref{2.14}) and (\ref{2.15}) are of type
MI-IV and hence $c_j \equiv \cn(x+(j-1)4K/p,m)$. Proceeding in the same way,
corresponding to most of the cyclic identities discussed in I and II,
 we can obtain the corresponding
cyclic identities with weighted terms.
For illustration purposes, a few
such cyclic identities are presented in Appendix F.

\section{Local Identities of Arbitrary Rank}

Now that we have obtained local and hence cyclic identities of low rank, 
the obvious question to ask is whether one
can generalize and obtain corresponding local (and hence also cyclic)
identities of arbitrary rank. In this section, we show that this is indeed possible. In particular, we show that corresponding to each local
low rank identity, we can obtain a local
identity of arbitrary rank
in which all the coefficients are explicitly
known.

As an illustration, let us start from the local identity
\beq\label{3.1}
\dn^2(x)[\dn(x+a)+\dn(x-a)]
= A\dn(x)+B[\dn(x+a)+\dn(x-a)]~,
\eeq
where $A,B$ are constants
\beq\label{3.2}
A= 2\ds(a)\ns(a)~,~~B = -\cs^2(a)~.
\eeq
This identity can be easily derived  using the addition formula
(\ref{2.1}). On repeatedly multiplying both sides of identity (\ref{3.1})
by $\dn^2(x)$ and using eq. (\ref{3.1}) we obtain the following
local identity of arbitrary odd rank
\beq\label{3.3}
\dn^{2n}(x)[\dn(x+a)+\dn(x-a)]
= A\sum_{k=1}^{n} B^{k-1} \dn^{2(n-k)+1}(x)
+B^{n}[\dn(x+a)+\dn(x-a)]~,
\eeq
where the constants $A,B$ are as given by eq. (\ref{3.2}).
Using the procedure described in Sec. 2, the corresponding
cyclic identities with and without
arbitrary weight $\omega$ are immediately obtained:
\beq\label{3.4}
\sum_{j=1}^{p} d_j^{2n}  [d_{j+r}+d_{j-r}]
= A\sum_{j=1}^{p}\,\sum_{k=1}^{n} B^{k-1} d_j^{2(n-k)+1}
+2B^{n} \sum_{j=1}^{p} d_j~,
\eeq
\beq\label{3.5}
\sum_{j=1}^{p} \omega^{j-1} d_j^{2n}
 [d_{j+r}+d_{j-r}]
= A\sum_{j=1}^{p}\,\sum_{k=1}^{n} \omega^{j-1} B^{k-1} d_j^{2(n-k)+1}
+2B^{n} \cos (\f{2\pi}{s})
\sum_{j=1}^{p} \omega^{j-1} d_j~,
\eeq
where $A,B$ and $\omega$ are as given by eqs. (\ref{3.2}) and (\ref{4})
respectively while $a=2rK/p$ and $p=0$ mod $s$.
Note that the identity without any weight (i.e. (\ref{3.4}))
is obtained from (\ref{3.5}) in the limit $s=1$ while for $s=2$ we obtain
the identity with alternate sign.
It is worth emphasizing that in the identities (\ref{3.3}) to (\ref{3.5})
of arbitrary odd rank, all the coefficients are explicitly known.

In order to obtain the corresponding local identity of arbitrary even rank,
we start from the identity (\ref{3.3}) and multiply both sides of it by
$\dn(x)$ and use the local identity (\ref{2.5}) to obtain
\beq\label{3.6}
\dn^{2n+1}(x) [\dn(x+a)+\dn(x-a)]
= A\sum_{k=1}^{n} B^{k-1} \dn^{2(n-k+1)}(x)
+2B^{n} \dn(a)
+B^{n}\cs(a)[\Z(x+a)-\Z(x-a) -2\Z(a)].
\eeq
The corresponding cyclic identities with and without arbitrary weight
are then immediately obtained by following the steps as above.

Proceeding in the same way, but starting from the identity
\beq\label{3.7}
\dn^2(x) [\dn(x+a)-\dn(x-a)]
= D\cn(x) \sn(x)+B[\dn(x+a)-\dn(x-a)]~,~~
D=-2\m\cs(a)~,
\eeq
multiplying recursively by $\dn^2(x)$ and using the identity (\ref{3.7}),
we obtain the following identities of arbitrary odd and even rank
\beq\label{3.8}
\dn^{2n}(x) [\dn(x+a)-\dn(x-a)]
= D\sum_{k=1}^{n} B^{k-1} \cn(x) \sn(x) \dn^{2(n-k)}(x)
+B^{n} [\dn(x+a)-\dn(x-a)]~,
\eeq
\beq\label{3.9}
\dn^{2n+1}(x) [\dn(x+a)-\dn(x-a)]
= D\sum_{k=1}^{n} B^{k-1} \cn(x) \sn(x) \dn^{2(n-k)+1}(x)
+B^{n} \cs(a)[\Z(x+a)+\Z(x-a)-2\Z(x) ],
\eeq
where $B=-\cs^2(a)$ and $D=-2\m\cs(a)$. The
corresponding cyclic identities are then immediately written down. Further,
by adding the two identities (\ref{3.3}) and (\ref{3.8}), we obtain the
basic local cyclic identity of any odd rank:
\beq\label{3.10}
\dn^{2n}(x) \dn(x+a)
= \f{D}{2} \sum_{k=1}^{n} B^{k-1} \cn(x) \sn(x) \dn^{2(n-k)}(x)
+B^{n} \dn(x+a)
+ \f{A}{2} \sum_{k=1}^{n} B^{k-1} \dn^{2(n-k)+1}(x)~,
\eeq
where $A,B,D$ are given by eqs. (\ref{3.2}) and (\ref{3.7}).
It is worth pointing out that using this identity, we can immediately
write down the local identity for the combination
$\dn(x) \dn^{2n}(x+a)$. This is done by replacing $x$ by $x-a$ followed by
changing $a$ to $-a$ in eq. (\ref{3.10}). In this way we obtain
\beq\label{3.11}
\!\!\dn(x) \dn^{2n}(x+a)
= \f{A}{2} \!\!\sum_{k=1}^{n}\!\! B^{k-1} \dn^{2(n-k)+1}(x+a)-\f{D}{2} \!\!\sum_{k=1}^{n}\!\! B^{k-1} \cn(x+a) \sn(x+a) \dn^{2(n-k)}(x+a)
+B^{n} \dn(x).
\eeq
Now $\dn(x) \dn^{2n}(x-a)$ can be immediately obtained from here by replacing
$a$ by $-a$. We find that
\beq\label{3.12}
\sum_{j=1}^{p} d_j^{2n} \left [d_{j+r} \pm d_{j-r} \right ]
=\pm \sum_{j=1}^{p} d_j \left [d^{2n}_{j+r} \pm d^{2n}_{j-r} \right ]~,
\eeq
\beq\label{3.13}
\sum_{j=1}^{p} (-1)^{j-1} d_j^{2n} \left [d_{j+r} \pm d_{j-r} \right ]
=\mp \sum_{j=1}^{p} (-1)^{j-1} d_j
\left [d^{2n}_{j+r} \pm d^{2n}_{j-r} \right ]~.
\eeq
In the next section we shall see that similar relations are in fact true in general
for any such combinations of Jacobi elliptic functions.

Proceeding in the same way, by starting from each of the lower rank
identities given in  Appendices A, B, C we can write down the corresponding
identities of arbitrary even as well as odd rank. Some illustrative examples 
with arbitrary even as well as odd powers of 
$\dn(x)$ (or $\sn(x)$ or $\cn(x)$)
are given in Appendix E.

\section{Master Local Identities}

In ref. II, we derived four master identities
from which all the cyclic identities could be derived as special
cases. In this section we show how a similar procedure works at the
level of the local identities, thereby systematizing the identities
and providing a unified framework for them. Besides, rather than using
the addition formulas for Jacobi elliptic functions, the master
identities (MI) provide an alternative way to derive the local, and
hence also cyclic, identities. We may note here the differences that
arise in the two approaches: addition formulas do not lead to
unevaluated constants in the form of integrals on the right hand side,
while the MI, in particular one of the four classes of MI do; on the
other hand MI reduces the right hand side maximally to standard forms,
while the addition formulas approach needs considerable algebraic manipulation 
to attain the simplest final form. In any
case, the two approaches are of course compatible and either of them
may be used. In this section and in Sec. 6, we will use $z$ instead of
$x$ as the variable to emphasize that there is no restriction to the
real numbers.

The classification in ref. I and in the above sections has been in
terms of the polynomial order of the elliptic functions appearing in
the left hand side, called the rank of the identities. In contrast,
the master identities first use symmetries to identify four
classes. Within each class, the identities are characterized by a
number which is the highest order of the singularities in the
fundamental domain of the left hand side. Thus, the analytic structure
of the functions appearing on the left hand side of any identity
determines the constants and the form of the functions that appear on
the right hand side. It is then quite easy to write symbolic
manipulation programs that turn any given form of the left hand side
into an appropriate local identity.

We first recall essential details of the analytic properties of
the Jacobi elliptic functions \cite{bf}. The function $\dn(z)$
is an even elliptic function of order two; there are two simple
poles inside the period parallelogram $(0,2K,2K+4iK',4iK')$
situated at $i K'$ and $3i K'$ with residues of $-i$ and $i$
respectively. The function $\sn(z)$ is an odd
elliptic function of order two; with two simple poles situated at
$i K'$ and $i K'+2K$, with residues $1/\sqrt{\m}$ and
$-1/\sqrt{\m}$, inside the fundamental period parallelogram
$(0,4K,4K+2iK',2iK')$. The function $\cn(z)$ is an even elliptic
function of order two; with two simple poles situated at $i K'$
and $2K +i K'$, with residues $-i/\sqrt{\m}$ and $i/\sqrt{\m}$,
inside the fundamental parallelogram $(-2K,2K,4K+2iK',2iK')$. We
note that the lattice of the poles in the complex plane is
identical for all these three functions. However these functions
have the following important distinguishing properties that we
will use below: $\dn(z+2iK')=-\dn(z)$, $\sn(z+2iK')=\sn(z)$,
$\cn(z+2iK')=-\cn(z)$, $\dn(z+2K)=\dn(z)$, $\sn(z+2K)=-\sn(z)$,
$\cn(z+2K)=-\cn(z)$.

The symmetry and periodicity properties put together allow us to
concentrate on the region $(0,2K,2K+2iK',2iK')$ uniformly for all
the functions, and consider only {\it one}  simple pole at $i
K'$. We supplement these possible symmetries with one additional
one, for which we describe the properties of the elliptic function
$\dn^2(z)$. Equivalently one may choose $\cn^2(z)$ or $\sn^2(z)$.
The function $\dn^2(z)$ has the fundamental domain
$(0,2K,2K+2iK',2iK')$ and consequently is completely periodic with
respect to translations of $2K$ and $2iK'$. It is also of order
two, with one double pole at $i K'$, with a residue of $0$. Thus
we classify functions $f(z)$ constructed from the Jacobian
elliptic functions into four symmetry classes. We define the
quantities $P,Q$ by: \beq f(z + 2i K')= (-1)^P f(z),\;\;
f(z+2K)=(-1)^Q f(z), \;\; P,Q=0,1~. \eeq

 We denote the four
possibilities by $(-,+)$, $(+,+)$, $(+,-)$ and $(-,-)$, where the
first sign refers to the sign of $(-1)^P$ and the second to that
of $(-1)^Q$. We note that as far as periodicity is concerned these
functions are identical to $\dn(z)$, $\dn^2(z)$, $\sn(z)$ and
$\cn(z)$ respectively. We also note that repeated differentiation
does not change the symmetry class to which the functions belong,
while this creates functions with arbitrarily high order of poles.
This then allows us to tailor suitable combinations of derivatives
of these four functions such that not only the periodicity, but
also the singular parts match with the given function $f(z)$. Thus
the difference between the function $f(z)$ and the tailored
combination is an elliptic function with no poles anywhere
including at infinity.
 We then use Liouville's theorem that states that if an analytic
function has no pole anywhere including at infinity then it must
be a constant, and then explicitly show that the constant is zero
in all cases except the second, where it can be evaluated as a
definite integral.

\subsection{Master Identities of Types I, III and IV}

Let $f(z)$ be an elliptic function with the symmetry properties
corresponding to $P=1,\,Q=0$, (type I)  and having $n_p$ poles
at positions $a_r$ $(r=1,\ldots,n_p)$ within the region
$(0,2K,2K+2iK',2iK')$ which we will call $ABCD$. Let the principal
part around the pole $a_r$ be \beq \sum_{l_r=1}^{L_r}
\f{\alpha_{l_r}^{(r)}}{(z-a_r)^{l_r}} \eeq
 We note that the principal part of $\dn(z)$ around the pole $i K'$ is \beq
\f{-i}{(z-i K')}.\eeq
Therefore if we consider the function $g(z)$:
\beq g(z)=\sum_{r=1}^{n_p} \sum_{l_r=1}^{L_r} \f{i
(-1)^{l_r-1}}{(l_r-1)!} \alpha^{(r)}_{l_r} \f{d^{l_r-1} \,
\dn(z)}{dz^{l_r-1}} |_{z-a_r+i K'} \eeq
this has identical poles as $f(z)$ and at these poles also has
identical principal parts. Due to the symmetry requirements, the
functions $f(z)$ and $g(z)$ also have identical periods and hence
they by Liouville's theorem they can differ utmost by a constant
that is independent of $z$. However integrating both these
functions from $0$ to $4iK'$, we see from the antisymmetry that
these must vanish, implying that the constant is zero; and hence
$f(z)=g(z)$. This is our ``master'' local identity of type I;
often the evaluated function $g(z)$ is of a simpler form than
$f(z)$.

Consider as an illustration the identity that results when $f(z)=
\dn^2(z)[\dn(z+a)+\dn(z-a)]$. This function has three poles,
within $ABCD$ at $a_1=i K'$, $a_2=-a+i K'$, and $a_3=a+i K'$. This
function has $P=1,\, Q=0$, and hence is of type I. At $a_1\equiv i
K' $ the principal part is
\[
-2 \,i \ds(a) \ns(a)/(z-i K').
\]
We note that although $\dn^2(z)$ has a double pole at $i K'$ it
gets ``softened'' by one, because $\dn(z+a)+\dn(z-a)$ has a zero
at $i K'$ for all $a$. This is the reason why we expect that the
RHS of such identities are simpler than the LHS. Thus
$\alpha^{(1)}_1=-2i\, \ds(a)\ns(a)$ and $L_1=1$. Similarly we get:
$\alpha^{(2)}_1=i \,\cs^2(a)$, $L_2=1$ and $\alpha^{(3)}_1=i\,
\cs^2(a)$, $L_3=1$. Hence this master identity yields the already
stated result in Eq.~(\ref{3.1}), which was alternatively derived
using addition formulas.

We note that in the case of cyclic identities further
simplification occurs and the $\alpha$ at the various poles can be
summed up, while in the case of local identities they are left as
they are. We note that the structure of the LHS yielded to
simplification  because roughly a zero cancelled a pole. We can
look at the ``parts'' of this identity where this does not happen
fully. Thus when we take $f(z)=\dn^2(z) \dn(z-a)$ we get the
identity:
\beq \label{halflocd2d} \dn^2(z) \dn(z-a)=\ds(a)\ns(a)\,
\dn(z)+m\, \cs(a) \, \cn(z) \sn(z) -\cs^2(a) \dn(z-a).\eeq
We note that the rank of the RHS is one less than that of the LHS,
and further reduction by one occurs when $a$ is changed to $-a$
and the two identities for $\dn^2(z) \dn(z+a)$ and $\dn^2(z) \dn(z-a)$
are added which results in the already
quoted identity of Eq.~(\ref{3.1}).

Similarly we derive the master identity for functions belonging to
type III $(P=0,Q=1)$:
\beq f(z)=\sum_{r=1}^{n_p} \sum_{l_r=1}^{L_r} \f{\sqrt{m}
(-1)^{l_r-1}}{(l_r-1)!} \alpha^{(r)}_{l_r} \f{d^{l_r-1} \,
\sn(z)}{dz^{l_r-1}} |_{z-a_r+i K'}, \eeq
and type IV $(P=1,Q=1)$:
\beq f(z)=\sum_{r=1}^{n_p} \sum_{l_r=1}^{L_r} \f{i \sqrt{m}
(-1)^{l_r-1}}{(l_r-1)!} \alpha^{(r)}_{l_r} \f{d^{l_r-1} \,
\cn(z)}{dz^{l_r-1}} |_{z-a_r+i K'}. \eeq
That the constant is zero in the case of type III and type IV
identities can be seen by integrating both sides from $0$ to $4K$.
It may be noted that symbolic manipulation packages that calculate
series expansions can be effectively used to generate these
identities.

\subsection{Master Identity of Type II}

This last type of identity deserves special mention; firstly the
function archetype is $\dn^2(z)$ which has a double pole at $i
K'$, secondly it leads to identities with non-zero constants, and
lastly the Jacobian zeta function appears in an essential way. A
function belonging to this type is periodic with periods $2K$ and
$2i K'$ (hence $P=0,Q=0$). Its principal part around $i K'$ is
\[ -1/(z-i K')^2. \]

Thus we write the master identity in this case as
\beq \label{master4} f(z)=C\,+\,\sum_{r=1}^{n_p}
\sum_{l_r=1}^{L_r} \f{ (-1)^{l_r-1}}{(l_r-1)!} \alpha^{(r)}_{l_r}
\f{d^{l_r-2} \, \dn^2(z)}{dz^{l_r-2}} |_{z-a_r+i K'}, \eeq
where $C$ is a constant. Note that the derivative order starts
from $-1$, which should be interpreted as an integral. Functions
of this type can have also simple poles and therefore the function
$\dn^2(z)$ and its derivatives are not sufficient to construct
these. If we include its integral the master identity is complete,
therefore we note the standard result \cite{bf} which can be taken
to be the definition of the Jacobian zeta function $Z(z)$:
\beq Z(z)=\int_{0}^{z} \left[ \dn^2(u)\,-\, \f{E}{K}\right] \, du
= E(z)-\f{E}{K}z,\eeq
where $E(z)$ is the incomplete elliptic integral of the second
kind and $E$ and $K$ are the complete elliptic integrals of the
second and first kinds respectively. Thus $Z(z)$ is closely
related to the incomplete elliptic integral of the second kind,
and is periodic with a period of $2 K$, but is not elliptic. It is
however {\it almost} elliptic due to the identity \cite{bf}:
$Z(z+2i K')=Z(z)-i \pi/K$. If we write the $l_r=1$ part of this
master equation, which has the only part with the Jacobian zeta
function, it is
\beq \sum_{r=1}^{n_p} \alpha^{(r)}_{1} \, Z(z-a_r+i K').\eeq
We note that {\it this} is an elliptic function with the correct
periods of $2K$ and $2i K'$, due to the fact that:
\beq \sum_{r=1}^{n_p} \alpha^{(r)}_{1}=0.\eeq
This is the sum of the {\it residues} of the function at all the
poles in $ABCD$. Making use of the double periodicity of $f(z)$ we
find the integral around $ABCD$ vanishes, hence from Cauchy's
theorem it follows that the sum of the residues must also vanish,
hence proving the above. Thus we are justified in using the zeta
function in this type of master identity even if it is not
elliptic: it will always appear in combinations that are elliptic
functions.

Integrating both sides of the master identity from $0$ to $2K$ we
get to evaluate the constant $C$:
\beq C=\f{1}{2K} \int_{0}^{2K} f(z)\, dz \, +\, \f{\gamma_2 E}{K},
\eeq where $\gamma_2=\sum_{r=1}^{n_p} \alpha^{(r)}_2$. We can
therefore evaluate the identity at some convenient $z$ where there
is no singularity (for instance perhaps $z=0$) and then make use
of the above to evaluate definite integrals. Assuming for instance
that there is no pole at $z=0$ we may write:
\beq \label{integralform}  \f{1}{2K}\int_{0}^{2K} f(z)\, dz=
f(0)\,-\,\gamma_2 \f{E}{K} \,- \,\sum_{r=1}^{n_p}
\sum_{l_r=1}^{L_r} \f{ (-1)^{l_r-1}}{(l_r-1)!} \alpha^{(r)}_{l_r}
\f{d^{l_r-2} \, \dn^2(z)}{dz^{l_r-2}} |_{i K'-a_r} \eeq
We note that this integral cannot be evaluated by a direct
application of Cauchy's theorem due to the vanishing of both the
contour integral around $ABCD$ and its residue. This is an useful
way of integrating many functions.

We also point out here something that is of relevance to the
cyclic identities:
\beq \sum_{j=1}^{p} g(z_j)[h(z_{j+1}) \pm h(z_{j-1})]=\pm
\sum_{j=1}^{p} h(z_j)[g(z_{j+1}) \pm g(z_{j-1})], \eeq
\beq \sum_{j=1}^{p} (-1)^j g(z_j)[h(z_{j+1}) \pm h(z_{j-1})]=\mp
\sum_{j=1}^{p} (-1)^j h(z_j)[g(z_{j+1}) \pm g(z_{j-1})], \eeq
where $h(z)$ and $g(z)$ are combinations of Jacobian elliptic
functions as above. These relate ordinary and alternating sums under
an interchange of $h$ and $g$ and follow from quite general
considerations related only to the periodicity of $h$ and $g$. The RHS
amounts to a rewriting of the LHS if we recall that either
$h(z_{p+1})= h(z_1)$ and $g(z_{p+1})= g(z_1)$ or $h(z_{p+1})= - h(z_1)$ 
and $g(z_{p+1})= - g(z_1)$ as the whole function $g(z)[h(z+T/p)+h(z-T/p)]$ is periodic with period $T$.

\section{Evaluation of Several Elliptic Integrals}

In ref. II, we obtained several cyclic identities of type MI-II in 
which the right hand side contained a definite 
integral involving products of Jacobi elliptic functions. These
integrals are not available  
in standard tables of integrals \cite{gr,bf}. We now
show that using local identities we can explicitly evaluate many such definite integrals.

As an illustration, we start from local identity (\ref{c15}).
On integrating both sides with
respect to $x$ over an interval $[0,2K]$ yields the definite
integral \beq\label{5.2} \int_0^{2K}\, \dn^2 (x) \dn^2 (x+a) \, dx
\,=\, -4E \cs^2(a)+2K[\cs^2(a)+\ds^2(a)-2\cs(a)\ds(a)\ns(a)
Z(a)]~. \eeq It may be noted that $a$ is here any non-zero
constant. 
Using this value of the integral in the MI-II cyclic identity (\ref{5}) that we obtained in ref. II and choosing $a=\f{2rK}{p}$, 
immediately yields the cyclic identity (\ref{2.10a}) 
which we had directly obtained
from the local identity. 

The other definite integrals which we are now able to evaluate are
related to cyclic identities containing an even number of $\dn$ or $\sn$
or $\cn$. For example, in ref. II, we simply stated the identity
\beq\label{5.3}
\frac{1}{p}\sum_{j=1}^{p} d_j d_{j+r} d_{j+s} d_{j+t} \equiv A
= \f{1}{2K} \int_{0}^{2K} \dn(x)\dn(x+a)\dn(x+a')
\dn(x+a'') \, dx~,
\eeq
but were unable to evaluate the integral and hence find $A$. Here,
$a=\f{2rK}{p}, a'=\f{2sK}{p}, a''=\f{2tK}{p}$. We now show
that using the local identities, we can in fact evaluate this definite integral
even for arbitrary but unequal $a, a', a''$.
To this purpose, we start from the local identity (\ref{dddd}). Integrating
both sides over the interval $[0,2K]$ yields 
\beqa\label{5.6}
&& \f{1}{2K} \int_{0}^{2K} \dn(x)\dn(x+a)\dn(x+a')
\dn(x+a'') \, dx =\dn(a)\dn(a')\dn(a'')+\cs(a)\cs(a'-a)\cs(a''-a)\Z(a) \nonumber \\
&&~~~~~~~~~~~~~~~~~~~-\cs(a')\cs(a'-a)\cs(a''-a')\Z(a')
+\cs(a'')\cs(a''-a)\cs(a''-a')\Z(a''),
\eeqa
which is eq. (\ref{g7}). Note that 
the special case $a=\f{2rK}{p}, a'=\f{2sK}{p}, a''=\f{2tK}{p}$
yields the cyclic identity (\ref{5.3}).

Finally, there are some MI-II cyclic identities and hence definite
integrals which were not even discussed in ref. II. For example, 
consider the local identity (\ref{dsc}) from which by following the method
explained in Sec. 2 we can deduce the cyclic identity
\beq\label{5.8}
\frac{1}{p} \sum_{j=1}^{p} \m d_j s_{j+r} c_{j+s} =
\frac{1}{2K} \int_{0}^{2K} \m \dn(x)\sn(x+a)\cn(x+a')
 \, dx, 
\eeq
where $a=\f{2rK}{p}, a'=\f{2sK}{p}$. 
By using the local identity (\ref{dsc}), 
we can in fact obtain this integral
for arbitrary but unequal values of $a, a'$. In particular,
on integrating both sides of
eq. (\ref{dsc}) over the interval $[0,2K]$ yields 
\beq\label{5.8a}
\f{1}{2K} \int_{0}^{2K} \!\!\m \dn(x)\sn(x+a)\cn(x+a')
 \, dx 
=-\ds(a-a')[\dn(a)-\cs(a)\Z(a)]
+\ns(a-a')[\dn(a')-\cs(a')\Z(a')]. 
\eeq
In the special case when $a=\f{2rK}{p}, a'=\f{2sK}{p},$ one recovers
the cyclic identity (\ref{5.8}).

Proceeding in the same way, we have been able to obtain expressions
for all the definite
integrals which appear in the cyclic identities of the type MI-II (and
which to the best of our knowledge are not known in the
literature). The answers for some of these definite integrals involving Jacobi elliptic functions are
given in Appendix G.

Actually, we can even simplify several indefinite integrals. 
In fact, by starting from any local identity
we can obtain the
indefinite integral of its left hand side in terms of the well known
integrals \cite{gr,bf}
of $\sn^{n} (x),\dn^{n} (x),\cn^{n} (x)$. The only exceptions are those MI-II
local identities in which the right hand side has a term proportional either
to $[\Z(x+a)+\Z(x-a)-2\Z(x)]$ or to
$[\Z(x+a)-\Z(x-a)]$. We show below that in that case the indefinite integral 
of the left hand side
also has terms containing indefinite elliptic integrals of the first, second
and third kind \cite{gr,bf}.

We start from the local identity
\beq\label{5.9}
\dn^2(x)\dn(x+a)=B\dn(x+a)+\ds(a)\ns(a)\dn(x)-\m \cs(a)\sn(x)\cn(x)~,~~
B \equiv -\cs^2(a)~,
\eeq
which is obtained by  adding the identities (\ref{3.1}) and (\ref{3.7}).
On integrating both sides with respect to $x$ and using the known integral
of $\dn(x)$ \cite{gr} we then obtain
\beq\label{5.10}
\int \, \dn^2(x)\dn(x+a) \, dx = B\am(x+a)+\ds(a) \ns(a) \am(x)+\cs(a)\dn(x)~,
\eeq
where $\int \dn(x) \, dx
=\am(x)=\sin^{-1}(\sn(x))=i\ln\left [\cn(x)-i\sn(x) \right ]$.
On multiplying both sides of eq. (\ref{5.9}) by $\dn^2(x)$ and using
eq. (\ref{5.10}) we then find that
\beqa\label{5.11}
&&\int \, \dn^4(x)\dn(x+a) \, dx = B^2\am(x+a) \nonumber \\
&&+\ds(a) \ns(a) \left [1+B-\f{\m}{2} \right ]\am(x)
+B\cs(a) \dn(x)+\f{\cs(a)}{3}\dn^3(x)+\f{\m\ds(a)\ns(a)}{2} \sn(x)\cn(x)~.
\eeqa

Proceeding in this way, one can find the indefinite integral of
$\dn^{2n}(x)\dn(x+a)$. One can also derive a recursion relation relating
the various integrals. In particular, using eq. (\ref{3.10})
one can show that
\beq\label{5.12}
\dn^{2n}(x)\dn(x+a)=B\dn^{2n-2}(x)\dn(x+a)+\ds(a) \ns(a) \dn^{2n-1}(x)
-\m \cs(a) \sn(x)\cn(x)\dn^{2n-2}(x)~.
\eeq
On integrating both sides with respect to $x$, we find a recursion
relation relating various integrals:
\beq\label{5.13}
I_{n}=BI_{n-1}+\f{\cs(a)}{(2n-1)} \dn^{2n-1}(x)
+\ds(a) \ns(a) \int \, \dn^{2n-1}(x) \, dx~;~~I_k \equiv \int \, \dn^{2k}(x)\dn(x+a) \, dx.
\eeq
Note that the integral of any power of $\dn(x)$ (as well
as $\sn(x),\cn(x)$) is known
in principle \cite{gr}.
We might also add that once the integral of say $\dn^{2n}(x) \dn(x+a)$ is known
then the integral of $\dn(x) \dn^{2n} (x+a)$ is obtained from it by simply
replacing $x$ by $x-a$ followed by $a \rightarrow -a$.

Using the above procedure one can obtain indefinite integrals of the left hand
sides of all the local identities given in this paper except for those
MI-II local identities where
the combination $[\Z(x+a)-\Z(x-a)]$ or $[\Z(x+a)+\Z(x-a)-2\Z(x)]$
also occurs on the right hand side. Let us now explain how to handle these
integrals.

We start from the local identity (\ref{2.5}). On using eq. (\ref{2.1}) it is
easily shown that
\beq\label{5.14}
\dn(x)[\dn(x+a)+\dn(x-a)] = \f{2\dn(a)[1-\m\sn^2(x)]}{1-\m\sn^2(a)\sn^2(x)}~.
\eeq
On integrating both sides of eq. (\ref{2.5}) over x, using eq. (\ref{5.14})
and well known integrals (see integrals 336.01 and 337.01 of (\cite{bf})) we
finally find that
\beqa\label{5.15}
&&\int \dn(x)[\dn(x+a)+\dn(x-a)]\, dx =
2[\dn(a)-\cs(a)\Z(a)] x +\cs(a)\int [\Z(x+a) -\Z(x-a)] \, dx \nonumber \\
&&\hspace{1in}=2\ds(a)\ns(a)F(\am~ x,k)-2\dn(a)\cs^2(a) \Pi (\am~ x,k^2\sn^2(a),k)~,
\eeqa
where $k^2 =\m$. Here $F(\am ~x,k)$ and $\Pi(\am ~x,k^2\sn^2(a),k)$ are indefinite
elliptic integrals of first and third kind respectively. Similarly, on using
eqs. (\ref{2.1}) and (\ref{2.8}) it is easy to show that
\beq\label{5.16}
\int\!\!\dn(x)[\dn(x+a)-\dn(x-a)]\, dx =
\cs(a)\!\!\int [\Z(x+a)+\Z(x-a)-2\Z(x)] \, dx
=\cs(a) \ln [1-\m\sn^2(a)\sn^2(x)],
\eeq
and hence
\beqa\label{5.17}
&&\int \dn(x)\dn(x+a)\, dx =
[\dn(a)-\cs(a)\Z(a)] x -\cs(a)\int [\Z(x+a) -\Z(x)] \, dx \nonumber \\
&&\!\!\!\!\!\!=\ds(a)\ns(a)F(\am ~x,k)-\dn(a)\cs^2(a) \Pi (\am ~x,k^2\sn^2(a),k)+
(1/2)\cs(a) \ln [1-\m\sn^2(a)\sn^2(x)].
\eeqa

We now show that using eq. (\ref{5.17})  one can obtain the
indefinite integral of the left hand side of any local MI-II identity. 
As an illustration,
consider the local identity (\ref{e10}). It is easy to show from here that
\beqa\label{5.18}
&&\!\!\!\!\!\!\!\!\!\!\!\!\dn^{2n}(x)\dn^2(x+a) = B \dn^{2n-2}(x)[\dn^2(x)+\dn^2(x+a)]-(1\!-\!\m)\dn^{2n-2}(x)
+2B^{n-1}\ds(a)\ns(a)\dn(x+a)\dn(x) \nonumber \\
&&+2\ds(a)\ns(a) \left [\ds(a)\ns(a)\dn(x)-\m\cs(a)\sn(x)\cn(x) \right] 
\sum_{k=1}^{n-1} B^{k-1} [\dn(x)]^{2(n-k)-1}~, 
\eeqa
where $B \equiv -\cs^2(a)$ and $n \ge 2$. On integrating both sides of this
equation, we get the recursion relation
\beqa\label{5.19}
&&I_n = B I_{n-1}+\int B\dn^{2n} (x) \, dx -(1-\m)\int \dn^{2n-2} (x) \, dx
+2B^{n-1}\ds(a)\ns(a)\int \dn(x+a)\dn(x) \,dx \nonumber \\
&&+\cs(a)\ds(a)\ns(a) \sum_{k=1}^{n-1} B^{k-1} \f{\dn^{2n-2k}(x)}{n-k}
+2\ds^2(a)\ns^2(a) \sum_{k=1}^{n-1} \!\int B^{k-1} \dn^{2(n-k)}(x) \, dx~,
\eeqa
where $I_{1}$ is easily obtained by using the integral (\ref{5.17}) and 
the local identity (\ref{c15}).
We find
\beq\label{5.20}
\!\!I_1 \!\equiv \!\!\int\! \!\dn^2(x)\dn^2(x+a)\: dx 
=\!\!-\cs^2(a)[E(\am ~x,k)+E(\am(x+a),k)]
-(1-\m)x+2\ds(a)\ns(a)\!\!\int\!\! \dn(x+a)\dn(x) \:dx,
\eeq
where $E(\am ~x,k)$ is the indefinite elliptic integral of the second kind.

Recursion relations for several indefinite elliptic
integrals are given in Appendix H, where the right hand side is in terms of Jacobi
elliptic functions, their integrals, and indefinite elliptic integrals
of the first, second and third kind.

\section{Continuum Limit of Local and Cyclic Identities}

We now study what happens to the local identities as $a
\rightarrow 0$. Although the identities are not valid at $a=0$,
this limit leads to well known nonlinear ordinary differential
equations satisfied by the Jacobian elliptic functions. Thus we
will see that the local identities may be viewed as {\it exact}
discretization of these differential equations. This provides our
justification for calling these identities ``local''. This is to
be contrasted with the cyclic identities that are exact
discretization of integral identities. Just as the differential
equation can be integrated, the local identities are simply summed
to produce the cyclic identities.

Take the simple type I identity:
\beq \label{localdiff} \dn^2(z)[ \dn(z+a)+\dn(z-a)]=2\ds(a)\ns(a)
\dn(z)-\cs^2(a)[ \dn(z+a)+\dn(z-a)]~. \eeq
Since $\cs^2(a)$ has a pole of order two at $a=0$, we expand to
second order in $a$:
\beq \dn(z+a)+\dn(z-a)=2\dn(z)+a^2 \f{d^2}{dz^2} \dn(z) +O(a^3)~.
\eeq
Using
\beq \lim_{a \rightarrow
0}\;[\ds(a)\ns(a)-\cs^2(a)]\,=\,1-\f{m}{2}~, \eeq
then leads to the limiting differential equation:
\beq \label{diffdn} 
(2-\m)y-\f{d^2y}{dz^2}=2y^3~,
\eeq 
with
$y=\dn(z)$~.
Of the many applications of such differential equations, the most
straightforward one would be the interpretation of this as
Newton's equation of motion for a particle in a one-dimensional
double well potential, with $z$ taking the role of time.

It is well known that finite difference versions of
such nonlinear one-dimensional problems tend to exhibit chaos, and
therefore the difference equation does not have analytical
solutions in terms of the Jacobi elliptic functions. However we
precisely achieve this when converting this differential equation
into a finite difference equation in the following manner:
\beq (2-m) y(z) -
\f{[y(z+\D)+y(z-\D)-2y(z)]}{\D^2}=y^2 (z) [y(z+\D)+y(z-\D)]~. 
\eeq 
Of course depending on the smallness of $\D$, the above discrete
equation will only be an approximation to the actual solution $\dn(z)$. 
However, replacing $2-m$ in the difference
scheme with $2[\ds(\D)\ns(\D)-\cs^2(\D)]$ and the $1/\D^2$ factor
multiplying the second difference with $\cs^2(\D)$  leads to an
{\em exact} difference scheme:
\beq 2[\ds(\D)\ns(\D)-\cs^2(\D)]y_n
-\cs^2(\D)[y_{n+1}+y_{n-1}-2y_{n}]=y_n^2[y_{n+1}+y_{n-1}]~,\eeq
where  $y(n)\equiv y(z+n\D)$, which is identical to the local
identity in eq.~(\ref{localdiff}) when we put $\D \equiv a$. This
is of course just the reverse of the small $a$ limit.

The local identity implies that $y_n=\dn(a\:n)$ is a solution of
this difference equation with the initial condition
$y_0=1,\,y_1=\dn(a)$. The modulus parameter is $m$ in all the
elliptic functions involved. This is analogous to the differential
equation eq. (\ref{diffdn}) whose solution is $y(z)=\dn(z)$, when
the initial conditions are $y(0)=1, y'(0)=0$, where the prime
denotes differentiation. It may be noted that the local identity
in eq.~(\ref{halflocd2d}) also has the same continuum limit as the
above, and that the difference equation is therefore genuinely
second order rather than first. All of the two-point local
identities limit to well known differential equations, or to those
that can be derived from these.

The cyclic identities limit to definite integrals, but we may, with
little effort, also generalize to indefinite integrals. By way of
illustration again consider the type I local identity in
eq.~(\ref{localdiff}).  Let $z_i=z_0+A\,(i-1)/p$, with
$i=1,\ldots,p$. Here $A$ is arbitrary and $p$ is an integer, $z_i$
are the sample points between $z_0$ and $z \equiv z_0+A$. Chaining
the local identities at these sample points together we get:
\beqa \label{noncyc1} \sum_{i=1}^{p}
\dn^2(z_i)[\dn(z_{i+1})+\dn(z_{i-1})]=[2
\ds(a)\ns(a)-\cs^2(a)]\sum_{i=1}^{p}\dn(z_i)\nonumber \\
-\cs^2(a)[\dn(z_{p+1})-\dn(z_1)+\dn(z_0)-\dn(z_p)]~, \eeqa
with $a=A/p$. This becomes a cyclic identity when $A=2K$ in which
case the ``end correction'' that is the last term in the above
equation vanishes due to the periodicity of $\dn(z)$. Multiply
both sides by $A/p$ and then take the $a \rightarrow 0$ (or
equivalently $p \rightarrow \infty)$ limit. The end corrections
are rewritten and evaluated as follows:
\beqa -\f{A}{p}\cs^2(a)[\dn(z_{p+1})-\dn(z_1)+\dn(z_0)-\dn(z_p)]
\rightarrow
-\f{p}{A}[\dn(z_{p+1})-\dn(z_p)+\dn(z_0)-\dn(z_1)]\\\nonumber
\rightarrow \m \,\sn(z)\cn(z)-\m\,\sn(z_0)\cn(z_0)~. \nonumber \eeqa
Here we have used that the derivative of $\dn(z)$ is $-m\,
\sn(z)\cn(z)$. Thus the entire identity in eq.~(\ref{noncyc1})
limits effectively to the indefinite integral:
\beq \int \dn^3(z) \, dz =\f{1}{2}(2-\m)\int \dn(z) \, dz 
+\f{\m}{2} \sn(z)\cn(z)~,
\eeq
which is a standard identity, for instance eq.~(314.03) of Byrd
and Friedman \cite{bf}. Thus the cyclic identities are exact
Riemann sums of integral identities while local identities are
exact discretization of differential equations.

In the case of three-point local identities (those which have the
elliptic functions evaluated at three points, $z$, $z+a$ and
$z+a'$) there are more than one limiting cases. The case $a'
\rightarrow a$ or $a' \rightarrow 0$ leads to two-point local
identities involving $z$ and $z+a$. For instance starting from the
local identity in eq.~(\ref{b20}) and taking the $a \rightarrow
a'$ limit or the $a' \rightarrow 0$ limit leads essentially to the
local two-point identity in eq.~(\ref{halflocd2d}).

\section{Comments and Discussion}

In this paper, we have proved a wide class of local identities,
using which we have obtained corresponding cyclic identities with
arbitrary weights. These results are easily extended in several
directions. For example, we could evaluate these identities at points
separated by gaps of $T/p$ with imaginary or complex period $T$
thereby obtaining corresponding local identities for pure imaginary as well
as complex shifts. Secondly, we can also obtain similar identities for all
9 auxiliary Jacobi functions like $\ns(x,\m)$ as well as 6 ratios of Jacobi functions
like $\cn(x)\dn(x) /\sn(x)$. Further, by following the procedure in ref. II, we can
readily write down the corresponding local as well as cyclic identities for
Weierstrass functions, as well as for the
ratio of any two of the four Jacobi theta functions.

Many of these identities have nontrivial $\m=0,1$ limits. For example, a 
cyclic identity derived in ref. II, which is valid for any odd integer $p$, and $l < p$, is
\beq\label{6.1}
\sum_{j=1}^{p} d_j d_{j+r}... d_{j+(l-1)r}
=\bigg [\prod_{k=1}^{(l-1)/2} \cs^2(ka)+2(-1)^{(l-1)/2} \sum_{k=1}^{(l-1)/2}
\prod_{n=1,n\ne k}^{l} \cs([n-k]a) \bigg ] \sum_{j=1}^{p} d_j~,
\eeq
while for $l=p$, we have the simpler identity
\beq\label{6.2}
\prod_{j=1}^{p} d_j
=\prod_{n=1}^{(p-1)/2} \cs^2 (\f{2Kn}{p}) \sum_{j=1}^{p} d_j~,
\eeq
where $a=r2K/p$. At $\m=0$, these identities reduce to interesting trigonometric
identities
\beq\label{6.3}
1=\prod_{k=1}^{(l-1)/2} \cot^2 (\f{rk\pi}{p})
+2(-1)^{(l-1)/2} \sum_{k=1}^{(l-1)/2}
\prod_{n=1,n\ne k}^{l} \cot([n-k]\f{r\pi}{p})~,
\eeq
\beq\label{6.4}
\frac{1}{p}=\prod_{n=1}^{(p-1)/2} \cot^2 (\f{n \pi}{p})~.
\eeq
Actually, identity (\ref{6.3}) is also valid for even $p$ and $r=1$,
provided $l < (p+2)/2$. 

For the special case of $l=3$, one can write down cyclic identities
for products of three $\dn$'s at arbitrary separation (in units of $2K/p$)
for both even and odd $p$. In particular, using the local identity (\ref{b20}),
one can immediately write down cyclic identities for combinations like
$\sum_{j=1}^{p} d_j d_{j+r} d_{j+s}$ with $r$ and $s$ being unequal but
arbitrary otherwise. In the limit $m=0$, summing of all such independent
cyclic identities yields the following remarkable trigonometric identities:
\beq\label{6o}
\f{(p-1)(p-2)}{3} = \sum_{j=1}^{p-1} \cot^2(\f{j\pi}{p})~.
\eeq
Similarly, many of the other local identities we have derived in this paper also reduce to interesting trigonometric identities in the limit $\m=0$. 

Finally, note that even though our local identities have been derived assuming
that $a$ is an arbitrary constant, the identities are also
valid when $a$ is any function of $x$. For example, on calling $x+a=b$,
the local identities (\ref{ds}) and (\ref{dc}) give generalized addition
theorems
\beq\label{6.8}
\dn(a-b) \sn(a) \sn(b)+\cn(a) \cn(b) = \cn(a-b)~,
\eeq
\beq\label{6.9}
\dn(a-b) \sn(a) \cn(b)-\cn(a) \sn(b) = -\dn(a) \sn(a-b)~,
\eeq
which in the limit $\m=0$ reduce to the well known addition theorems for the
trigonometric functions.

AK and US gratefully acknowledge support from the U.S. Department
of Energy.

\newpage

\noindent{\bf\Large Appendix A: Local Identities of Rank 2}

\sss\noindent{\bf\large Rank 2 identities with 2 distinct arguments ($x, x+a$).}

\beq \label{dd}
\dn(x)\dn(x+a)
=\dn(a)+\cs(a) \left [\Z(x+a)-\Z(x)-\Z(a)  \right ]~,
\eeq
\beq\label{a2}
\m\sn(x)\sn(x+a)
=-\ns(a)\left [\Z(x+a)-\Z(x)-\Z(a) \right ]~,
\eeq
\beq\label{a3}
\m\cn(x)\cn(x+a)
=\m\cn(a)+\ds(a)\left [\Z(x+a)-\Z(x)-\Z(a) \right ]~,
\eeq
\beq \label{ds}
\dn(x)\sn(x+a)
=\ns(a)\cn(x)-\cs(a)\cn(x+a)~,
\eeq
\beq \label{dc}
\dn(x)\cn(x+a)
=-\ds(a)\sn(x)+\cs(a)\sn(x+a)~,
\eeq
\beq \label{sc}
\m\sn(x)\cn(x+a)
=\ds(a)\dn(x)-\ns(a)\dn(x+a)~.
\eeq

\sss
\noindent{\bf\Large Appendix B: Local Identities of Rank 3}

\sss\noindent{\bf\large Rank 3 identities with 3 distinct arguments ($x, x+a, x+a'$).}

\beq \label{b20}
\dn(x)\dn(x+a)\dn(x+a')
=-\cs(a)\cs(a')\dn(x)-\cs(a)\cs(a-a')\dn(x+a)+\cs(a')\cs(a-a')\dn(x+a')~,
\eeq
\beq \label{sss}
\m\sn(x)\sn(x+a)\sn(x+a')
=\ns(a)\ns(a')\sn(x)+\ns(a)\ns(a-a')\sn(x+a)-\ns(a')\ns(a-a')\sn(x+a')~,
\eeq
\beq \label{ccc}
\m\cn(x)\cn(x+a)\cn(x+a')
=-\ds(a)\ds(a')\cn(x)-\ds(a)\ds(a-a')\cn(x+a)+\ds(a')\ds(a-a')\cn(x+a')~,
\eeq
\beq \label{dds}
\dn(x)\dn(x+a)\sn(x+a')
=-\cs(a)\ns(a')\sn(x)-\cs(a)\ns(a-a')\sn(x+a)+\cs(a')\cs(a-a')\sn(x+a')~,
\eeq
\beq \label{ddc}
\dn(x)\dn(x+a)\cn(x+a')
=-\cs(a)\ds(a')\cn(x)-\cs(a)\ds(a-a')\cn(x+a)+\cs(a')\cs(a-a')\cn(x+a')~,
\eeq
\beq \label{ssd}
\m\sn(x)\sn(x+a)\dn(x+a')
=\ns(a)\cs(a')\dn(x)+\ns(a)\cs(a-a')\dn(x+a)-\ns(a')\ns(a-a')\dn(x+a')~,
\eeq
\beq \label{ssc}
\m\sn(x)\sn(x+a)\cn(x+a')
=\ns(a)\ds(a')\cn(x)+\ns(a)\ds(a-a')\cn(x+a)-\ns(a')\ns(a-a')\cn(x+a')~,
\eeq
\beq \label{ccd}
\m\cn(x)\cn(x+a)\dn(x+a')
=-\ds(a)\cs(a')\dn(x)-\ds(a)\cs(a-a')\dn(x+a)+\ds(a')\ds(a-a')\dn(x+a')~,
\eeq
\beq \label{ccs}
\m\cn(x)\cn(x+a)\sn(x+a')
=-\ds(a)\ns(a')\sn(x)-\ds(a)\ns(a-a')\sn(x+a)+\ds(a')\ds(a-a')\sn(x+a')~,
\eeq
\beqa \label{dsc}
&&\m\dn(x)\sn(x+a)\cn(x+a')
=-\ds(a-a')\{\dn(a)+\cs(a)\left [\Z(x+a) -\Z(x) -\Z(a) \right ]
\} \nonumber \\
&&+\ns(a-a') \{\dn(a')+\cs(a') \left [\Z(x+a') -\Z(x) -\Z(a') \right ]
\}~.
\eeqa

\sss\noindent{\bf\large Rank 3 identities with 2 distinct arguments ($x, x+a$).}

\sss
\noindent As explained in the text,
local identities for say $\m\,\cn(x)\sn(x+a)\dn(x+a)$ and
$\m\,\sn(x)\dn(x)\cn(x+a)$ are related to each other by $x \rightarrow x-a$
followed by $a \rightarrow -a$ and hence only one of these identities is
given below.

\beq
\dn^2(x)\dn(x+a)
=-\cs^2(a)\dn(x+a)+\ds(a)\ns(a)\dn(x)-\m\cs(a)\cn(x)\sn(x)~,
\eeq
\beq
\m\sn^2(x)\sn(x+a)
=\ns^2(a)\sn(x+a)-\cs(a)\ds(a)\sn(x)-\ns(a)\cn(x)\dn(x)~,
\eeq
\beq
\m\cn^2(x)\cn(x+a)
=-\ds^2(a)\cn(x+a)+\cs(a)\ns(a)\cn(x)-\ds(a)\sn(x)\dn(x)~,
\eeq
\beq
\dn(x)\sn(x)\dn(x+a)
=-\cs(a)\ns(a)\sn(x+a)+\ds(a)\ns(a)\sn(x)+\cs(a)\cn(x)\dn(x)~,
\eeq
\beq
\dn(x)\cn(x)\dn(x+a)
=-\cs(a)\ds(a)\cn(x+a)+\ds(a)\ns(a)\cn(x)-\cs(a)\sn(x)\dn(x)~,
\eeq
\beq\
\m\dn(x)\sn(x)\sn(x+a)
=\cs(a)\ns(a)\dn(x+a)-\cs(a)\ds(a)\dn(x)+\m\ns(a)\cn(x)\sn(x)~,
\eeq
\beq
\m\sn(x)\cn(x)\sn(x+a)
=\ds(a)\ns(a)\cn(x+a)-\cs(a)\ds(a)\cn(x)+\ns(a)\sn(x)\dn(x)~,
\eeq
\beq
\m\dn(x)\cn(x)\cn(x+a)
=-\cs(a)\ds(a)\dn(x+a)+\cs(a)\ns(a)\dn(x)-\m\ds(a)\cn(x)\sn(x)~,
\eeq
\beq
\m\sn(x)\cn(x)\cn(x+a)
=-\ds(a)\ns(a)\sn(x+a)+\cs(a)\ns(a)\sn(x)+\ds(a)\cn(x)\dn(x)~,
\eeq
\beq
\m\dn(x)\sn(x)\cn(x+a)
=-\ds(a)-\cs(a)\ns(a) \left [\Z(x+a)-\Z(x)-\Z(a)  \right ]
+\ds(a)\dn^2(x)~,
\eeq
\beq
\m\cn(x)\dn(x)\sn(x+a)
=-\ds(a)\dn(a)-\cs(a)\ds(a) \left [\Z(x+a)-\Z(x)-\Z(a)  \right ]
+\ns(a)\dn^2(x)~,
\eeq
\beq
\m\sn(x)\cn(x)\dn(x+a)
=-\cs(a)-\ds(a)\ns(a) \left [\Z(x+a)-\Z(x)-\Z(a)  \right ]
+\cs(a)\dn^2(x)~.
\eeq

\sss
\noindent{\bf\Large Appendix C: Local Identities of Rank 4}

\sss\noindent{\bf\large Rank 4 identities with 4 distinct arguments ($x, x+a, x+a', x+a''$).}
\beqa\label{dddd}
&&\dn(x)\dn(x+a)\dn(x+a')\dn(x+a'')
=-\cs(a)\cs(a')\{ \dn(a'')+\cs(a'')\big[\Z(x+a'')-\Z(x)-\Z(a'')\big]
\} \nonumber \\
&&\hspace{.8in}-\cs(a)\cs(a-a')\{\dn(a''-a)+\cs(a''-a)
\big[\Z(x+a'')-\Z(x+a)-\Z(a''-a)\big] \} \nonumber \\
&&\hspace{.8in}+\cs(a')\cs(a-a')\{\dn(a''-a')+\cs(a''-a')
\big[\Z(x+a'')-\Z(x+a')-\Z(a''-a')\big] \}~,
\eeqa
\beqa\label{ssss}
&&\m^2\sn(x)\sn(x+a)\sn(x+a')\sn(x+a'')
=-\ns(a)\ns(a')\ns(a'')\big[\Z(x+a'')-\Z(x)-\Z(a'')\big]
 \nonumber \\
&&\hspace{1.2in}-\ns(a)\ns(a-a')\ns(a''-a)
\big[\Z(x+a'')-\Z(x+a)-\Z(a''-a)\big] \nonumber \\ 
&&\hspace{1.2in}+\ns(a')\ns(a-a')\ns(a''-a')
\big[\Z(x+a'')-\Z(x+a')-\Z(a''-a')\big]~,
\eeqa
\beqa\label{cccc}
&&\m^2\cn(x)\cn(x+a)\cn(x+a')\cn(x+a'')
=-\ds(a)\ds(a')\{\m\cn(a'')+\ds(a'')\big[\Z(x+a'')-\Z(x)-\Z(a'')\big]
\} \nonumber \\
&&\hspace{.8in}-\ds(a)\ds(a-a')\{\m\cn(a''-a)+\ds(a''-a)
\big[\Z(x+a'')-\Z(x+a)-\Z(a''-a)\big] \} \nonumber \\
&&\hspace{.8in}+\ds(a')\ds(a-a')\{\m\cn(a''-a')+\ds(a''-a')
\big[\Z(x+a'')-\Z(x+a')-\Z(a''-a')\big] \}~,
\eeqa
\beqa\label{ddds}
&&\dn(x)\dn(x+a)\dn(x+a')\sn(x+a'')
=-\cs(a)\cs(a')\big[\ns(a'')\cn(x)-\cs(a'')\cn(x+a'')\big] \nonumber \\
&&\hspace{1in}-\cs(a)\cs(a-a')
\big[\ns(a''-a)\cn(x+a)-\cs(a''-a)\cn(x+a'')\big] \nonumber \\ 
&&\hspace{1in}+\cs(a')\cs(a-a')
\big[\ns(a''-a')\cn(x+a')-\cs(a''-a')\cn(x+a'')\big]~, 
\eeqa
\beqa\label{dddc}
&&\dn(x)\dn(x+a)\dn(x+a')\cn(x+a'')
=\cs(a)\cs(a')\big[\ds(a'')\sn(x)-\cs(a'')\sn(x+a'')\big] \nonumber \\
&&\hspace{1in}+\cs(a)\cs(a-a')
\big[\ds(a''-a)\sn(x+a)-\cs(a''-a)\sn(x+a'')\big] \nonumber \\
&&\hspace{1in}-\cs(a')\cs(a-a')
\big[\ds(a''-a')\sn(x+a')-\cs(a''-a')\sn(x+a'')\big]~, 
\eeqa
\beqa\label{sssd}
&&\m\sn(x)\sn(x+a)\sn(x+a')\dn(x+a'')
=\ns(a)\ns(a')\big[\cs(a'')\cn(x)-\ns(a'')\cn(x+a'')\big] \nonumber \\
&&\hspace{1in}+\ns(a)\ns(a-a')
\big[\cs(a''-a)\cn(x+a)-\ns(a''-a)\cn(x+a'')\big] \nonumber \\ 
&&\hspace{1in}-\ns(a')\ns(a-a')
\big[\cs(a''-a')\cn(x+a')-\ns(a''-a')\cn(x+a'')\big]~, 
\eeqa
\beqa\label{sssc}
&&\m^2\sn(x)\sn(x+a)\sn(x+a')\cn(x+a'')
=\ns(a)\ns(a')\big[\ds(a'')\dn(x)-\ns(a'')\dn(x+a'')\big] \nonumber \\
&&\hspace{1in}+\ns(a)\ns(a-a')
\big[\ds(a''-a)\dn(x+a)-\ns(a''-a)\dn(x+a'')\big] \nonumber \\ 
&&\hspace{1in}-\ns(a')\ns(a-a')
\big[\ds(a''-a')\dn(x+a')-\ns(a''-a')\dn(x+a'')\big]~, 
\eeqa
\beqa\label{cccd}
&&\m\cn(x)\cn(x+a)\cn(x+a')\dn(x+a'')
=\ds(a)\ds(a')\big[\cs(a'')\sn(x)-\ds(a'')\sn(x+a'')\big] \nonumber \\
&&\hspace{1in}+\ds(a)\ds(a-a')
\big[\cs(a''-a)\sn(x+a)-\ds(a''-a)\sn(x+a'')\big] \nonumber \\ 
&&\hspace{1in}-\ds(a')\ds(a-a')
\big[\cs(a''-a')\sn(x+a')-\ds(a''-a')\sn(x+a'')\big]~, 
\eeqa
\beqa\label{cccs}
&&\m^2\cn(x)\cn(x+a)\cn(x+a')\sn(x+a'')
=-\ds(a)\ds(a')\big[\ns(a'')\dn(x)-\ds(a'')\dn(x+a'')\big] \nonumber \\
&&\hspace{1in}-\ds(a)\ds(a-a')
\big[\ns(a''-a)\dn(x+a)-\ds(a''-a)\dn(x+a'')\big] \nonumber \\
&&\hspace{1in}+\ds(a')\ds(a-a')
\big[\ns(a''-a')\dn(x+a')-\ds(a''-a')\dn(x+a'')\big]~, 
\eeqa
\beqa\label{ddss}
&&\m\sn(x)\dn(x+a)\sn(x+a')\dn(x+a'')
=\cs(a)\ns(a')\{\dn(a'')+\cs(a'')\big[\Z(x+a'')-\Z(x)-\Z(a'')\big]
\} \nonumber \\
&&\hspace{.8in}+\ns(a)\ns(a-a')\{\dn(a''-a)+\cs(a''-a)
\big[\Z(x+a'')-\Z(x+a)-\Z(a''-a)\big] \} \nonumber \\
&&\hspace{.8in}-\ns(a')\cs(a-a')\{\dn(a''-a')+\cs(a''-a')
\big[\Z(x+a'')-\Z(x+a')-\Z(a''-a')\big] \}~,
\eeqa
\beqa\label{ddcc}
&&\m\cn(x)\dn(x+a)\cn(x+a')\dn(x+a'')
=-\cs(a)\ds(a')\{\dn(a'')+\cs(a'')\big[\Z(x+a'')-\Z(x)-\Z(a'')\big]
\} \nonumber \\
&&\hspace{.8in}-\ds(a)\ds(a-a')\{\dn(a''-a)+\cs(a''-a)
\big[\Z(x+a'')-\Z(x+a)-\Z(a''-a)\big] \} \nonumber \\
&&\hspace{.8in}+\ds(a')\cs(a-a')\{\dn(a''-a')+\cs(a''-a')
\big[\Z(x+a'')-\Z(x+a')-\Z(a''-a')\big] \}~,
\eeqa
\beqa\label{sscc}
&&\m^2\sn(x)\cn(x+a)\cn(x+a')\sn(x+a'')
=\ds(a)\ds(a')\ns(a'')\big[\Z(x+a'')-\Z(x)-\Z(a'')\big]
 \nonumber \\
&&\hspace{1in}+\ns(a)\ds(a-a')\ns(a''-a)
\big[\Z(x+a'')-\Z(x+a)-\Z(a''-a)\big] \nonumber \\  
&&\hspace{1in}-\ns(a')\ds(a-a')\ns(a''-a')
\big[\Z(x+a'')-\Z(x+a')-\Z(a''-a')\big]~,
\eeqa
\beqa\label{ddsc}
&&\m\cn(x)\dn(x+a)\dn(x+a')\sn(x+a'')
=-\cs(a)\cs(a')\big[\ns(a'')\dn(x)-\ds(a'')\dn(x+a'')\big] \nonumber \\
&&\hspace{1in}-\ds(a)\cs(a-a')
\big[\ns(a''-a)\dn(x+a)-\ds(a''-a)\dn(x+a'')\big] \nonumber \\
&&\hspace{1in}+\ds(a')\cs(a-a')
\big[\ns(a''-a')\dn(x+a')-\ds(a''-a')\dn(x+a'')\big]~, 
\eeqa
\beqa\label{ssdc}
&&\m\sn(x)\dn(x+a)\sn(x+a')\cn(x+a'')
=-\cs(a)\ns(a')\big[\ds(a'')\sn(x)-\cs(a'')\sn(x+a'')\big] \nonumber \\
&&\hspace{1in}-\ns(a)\ns(a-a')
\big[\ds(a''-a)\sn(x+a)-\cs(a''-a)\sn(x+a'')\big] \nonumber \\ 
&&\hspace{1in}+\ns(a')\cs(a-a')
\big[\ds(a''-a')\sn(x+a')-\cs(a''-a')\sn(x+a'')\big]~, 
\eeqa
\beqa\label{ccds}
&&\m\cn(x)\dn(x+a)\cn(x+a')\sn(x+a'')
=-\cs(a)\ds(a')\big[\ns(a'')\cn(x)-\cs(a'')\cn(x+a'')\big] \nonumber \\
&&\hspace{1in}-\ds(a)\ds(a-a')
\big[\ns(a''-a)\cn(x+a)-\cs(a''-a)\cn(x+a'')\big] \nonumber \\
&&\hspace{1in}+\ds(a')\cs(a-a')
\big[\ns(a''-a')\cn(x+a')-\cs(a''-a')\cn(x+a'')\big]~. 
\eeqa

\sss\noindent{\bf\large Rank 4 identities with 3 distinct arguments ($x, x+a, x+a'$).}

\beqa\label{c46}
&&\dn^2(x)\dn(x+a)\dn(x+a')
=-\cs(a)\cs(a-a')\{\dn(a)+\cs(a)
\big[\Z(x+a)-\Z(x)-\Z(a)\big] \} \nonumber \\
&&\hspace{.8in}+\cs(a')\cs(a-a')\{\dn(a')+\cs(a')
\big[\Z(x+a')-\Z(x)-\Z(a')\big] \} -\cs(a)\cs(a')\dn^2(x)~,
\eeqa
\beqa\label{c47}
&&\m^2\sn^2(x)\sn(x+a)\sn(x+a')
=-\ns^2(a)\ns(a-a')
\big[\Z(x+a)-\Z(x)-\Z(a)\big] \nonumber \\
&&\hspace{1in}+\ns^2(a')\ns(a-a')
\big[\Z(x+a')-\Z(x)-\Z(a')\big] +\m \ns(a)\ns(a')\sn^2(x)~,
\eeqa
\beqa\label{c48}
&&\m^2\cn^2(x)\cn(x+a)\cn(x+a')
=-\ds(a)\ds(a-a')\{\m\cn(a)+\ds(a)
\big[\Z(x+a)-\Z(x)-\Z(a)\big] \} \nonumber \\
&&\hspace{.4in}+\ds(a')\ds(a-a')\{\m\cn(a')+\ds(a')
\big[\Z(x+a')-\Z(x)-\Z(a')\big] \} -\m\ds(a)\ds(a')\cn^2(x)~,
\eeqa
\beqa\label{c49}
&&\m\dn(x)\sn(x)\dn(x+a)\sn(x+a')
=\cs(a)\ns(a)\ns(a-a')
\big[\Z(x+a)-\Z(x)-\Z(a)\big] \nonumber \\
&&\hspace{.8in}-\cs(a')\ns(a')\cs(a-a')
\big[\Z(x+a')-\Z(x)-\Z(a')\big] -\m \cs(a)\ns(a')\sn^2(x)~,
\eeqa
\beqa\label{c50}
&&\m\dn(x)\cn(x)\dn(x+a)\cn(x+a')
=-\cs(a)\ds(a-a')\{\m\cn(a)+\ds(a)
\big[\Z(x+a)-\Z(x)-\Z(a)\big] \} \nonumber \\
&&\hspace{.4in}+\cs(a')\cs(a-a')\{\m\cn(a')+\ds(a')
\big[\Z(x+a')-\Z(x)-\Z(a')\big] \} -\m\cs(a)\ds(a')\cn^2(x)~,
\eeqa
\beqa\label{c51}
&&\m^2\sn(x)\cn(x)\sn(x+a)\cn(x+a')
=\ds(a)\ns(a)\ds(a-a')
\big[\Z(x+a)-\Z(x)-\Z(a)\big] \nonumber \\
&&\hspace{.4in}-\ds(a')\ns(a')\ns(a-a')
\big[\Z(x+a')-\Z(x)-\Z(a')\big] -\m \ns(a)\ds(a')\sn^2(x)~,
\eeqa
\beqa\label{c52}
&&\dn(x)^2\dn(x+a)\sn(x+a')
=-\cs(a)\ns(a-a')
\big[\ns(a)\cn(x)-\cs(a)\cn(x+a) \big] \nonumber \\
&&\hspace{.4in}+\cs(a')\cs(a-a')
\big[\ns(a')\cn(x)-\cs(a')\cn(x+a') \big] 
-\cs(a)\ns(a')\sn(x)\dn(x)~,
\eeqa
\beqa\label{c53}
&&\dn(x)^2\dn(x+a)\cn(x+a')
=\cs(a)\ds(a-a')
\big[\ds(a)\sn(x)-\cs(a)\sn(x+a) \big] \nonumber \\
&&\hspace{.4in}-\cs(a')\cs(a-a')
\big[\ds(a')\sn(x)-\cs(a')\sn(x+a') \big] 
-\cs(a)\ds(a')\cn(x)\dn(x)~,
\eeqa
\beqa\label{c54}
&&\m^2\cn(x)^2\sn(x+a)\cn(x+a')
=-\ds(a)\ds(a-a')
\big[\ns(a)\dn(x)-\ds(a)\dn(x+a) \big] \nonumber \\
&&\hspace{.4in}+\ds(a')\ns(a-a')
\big[\ns(a')\dn(x)-\ds(a')\dn(x+a') \big] 
-\m\ns(a)\ds(a')\cn(x)\sn(x)~,
\eeqa
\beqa\label{c55}
&&\m\dn(x)\sn(x)\sn(x+a)\cn(x+a')
=-\ns(a)\ds(a-a')
\big[\ds(a)\sn(x)-\cs(a)\sn(x+a) \big] \nonumber \\
&&\hspace{.4in}+\ns(a')\ns(a-a')
\big[\ds(a')\sn(x)-\cs(a')\sn(x+a') \big] 
+\ns(a)\ds(a')\cn(x)\dn(x)~,
\eeqa
\beqa\label{c56}
&&\m\dn(x)\sn(x)\dn(x+a)\cn(x+a')
=-\cs(a)\ds(a-a')
\big[\ds(a)\dn(x)-\ns(a)\dn(x+a) \big] \nonumber \\
&&\hspace{.4in}+\cs(a')\cs(a-a')
\big[\ds(a')\dn(x)-\ns(a')\dn(x+a') \big] 
-\m\cs(a)\ds(a')\cn(x)\sn(x)~,
\eeqa
\beqa\label{c57}
&&\dn(x)\sn(x)\dn(x+a)\dn(x+a')
=-\cs(a)\cs(a-a')
\big[\cs(a)\cn(x)-\ns(a)\cn(x+a) \big] \nonumber \\
&&\hspace{.4in}+\cs(a')\cs(a-a')
\big[\cs(a')\cn(x)-\ns(a')\cn(x+a') \big] 
-\cs(a)\cs(a')\sn(x)\dn(x)~,
\eeqa
\beqa\label{c58}
&&\m\dn(x)\sn(x)\sn(x+a)\sn(x+a')
=\ns(a)\ns(a-a')
\big[\ns(a)\cn(x)-\cs(a)\cn(x+a) \big] \nonumber \\
&&\hspace{.4in}-\ns(a')\ns(a-a')
\big[\ns(a')\cn(x)-\cs(a')\cn(x+a') \big] 
+\ns(a)\ns(a')\sn(x)\dn(x)~,
\eeqa
\beqa\label{c59}
&&\m\dn(x)\sn(x)\cn(x+a)\cn(x+a')
=-\ns(a)\ds(a-a')
\big[\ns(a)\cn(x)-\cs(a)\cn(x+a) \big] \nonumber \\
&&\hspace{.4in}+\ns(a')\ds(a-a')
\big[\ns(a')\cn(x)-\cs(a')\cn(x+a') \big] 
-\ds(a)\ds(a')\sn(x)\dn(x)~,
\eeqa
\beqa\label{c60}
&&\m\dn(x)\cn(x)\sn(x+a)\cn(x+a')
=-\ds(a)\ds(a-a')
\big[\ns(a)\cn(x)-\cs(a)\cn(x+a) \big] \nonumber \\
&&\hspace{.4in}+\ds(a')\ns(a-a')
\big[\ns(a')\cn(x)-\cs(a')\cn(x+a') \big] 
-\ns(a)\ds(a')\sn(x)\dn(x)~,
\eeqa
\beqa\label{c61}
&&\m\dn(x)\cn(x)\dn(x+a)\sn(x+a')
=-\cs(a)\ns(a-a')
\big[\ns(a)\dn(x)-\ds(a)\dn(x+a) \big] \nonumber \\
&&\hspace{.4in}+\cs(a')\cs(a-a')
\big[\ns(a')\dn(x)-\ds(a')\dn(x+a') \big] 
-\m\cs(a)\ns(a')\cn(x)\sn(x)~,
\eeqa
\beqa\label{c62}
&&\dn(x)\cn(x)\dn(x+a)\dn(x+a')
=\cs(a)\cs(a-a')
\big[\cs(a)\sn(x)-\ds(a)\sn(x+a) \big] \nonumber \\
&&\hspace{.4in}-\cs(a')\cs(a-a')
\big[\cs(a')\sn(x)-\ds(a')\sn(x+a') \big] 
-\cs(a)\cs(a')\cn(x)\dn(x)~,
\eeqa
\beqa\label{c63}
&&\m\dn(x)\cn(x)\sn(x+a)\sn(x+a')
=-\ds(a)\ns(a-a')
\big[\ds(a)\sn(x)-\cs(a)\sn(x+a) \big] \nonumber \\
&&\hspace{.4in}+\ds(a')\ns(a-a')
\big[\ds(a')\sn(x)-\cs(a')\sn(x+a') \big] 
+\ns(a)\ns(a')\cn(x)\dn(x)~,
\eeqa
\beqa\label{c64}
&&\m\dn(x)\cn(x)\cn(x+a)\cn(x+a')
=\ds(a)\ds(a-a')
\big[\ds(a)\sn(x)-\cs(a)\sn(x+a) \big] \nonumber \\
&&\hspace{.4in}-\ds(a')\ds(a-a')
\big[\ds(a')\sn(x)-\cs(a')\sn(x+a') \big] 
-\ds(a)\ds(a')\cn(x)\dn(x)~,
\eeqa
\beqa\label{c65}
&&\m^2\sn(x)\cn(x)\sn(x+a)\sn(x+a')
=\ds(a)\ns(a-a')
\big[\ds(a)\dn(x)-\ns(a)\dn(x+a) \big] \nonumber \\
&&\hspace{.4in}-\ds(a')\ns(a-a')
\big[\ds(a')\dn(x)-\ns(a')\dn(x+a') \big] 
+\m\ns(a)\ns(a')\sn(x)\cn(x)~,
\eeqa
\beqa\label{c66}
&&\m\sn(x)\cn(x)\dn(x+a)\dn(x+a')
=-\ns(a)\cs(a-a')
\big[\ns(a)\dn(x)-\ds(a)\dn(x+a) \big] \nonumber \\
&&\hspace{.4in}+\ns(a')\cs(a-a')
\big[\ns(a')\dn(x)-\ds(a')\dn(x+a') \big] 
-\m\cs(a)\cs(a')\cn(x)\sn(x)~,
\eeqa
\beqa\label{c67}
&&\m^2\sn(x)\cn(x)\cn(x+a)\cn(x+a')
=-\ns(a)\ds(a-a')
\big[\ns(a)\dn(x)-\ds(a)\dn(x+a) \big] \nonumber \\
&&\hspace{.4in}+\ns(a')\ds(a-a')
\big[\ns(a')\dn(x)-\ds(a')\dn(x+a') \big] 
-\m\ds(a)\ds(a')\cn(x)\sn(x)~,
\eeqa
\beqa\label{c68}
&&\m\sn(x)\cn(x)\dn(x+a)\sn(x+a')
=-\ns(a)\ns(a-a')
\big[\cs(a)\sn(x)-\ds(a)\sn(x+a) \big] \nonumber \\
&&\hspace{.4in}+\ns(a')\cs(a-a')
\big[\cs(a')\sn(x)-\ds(a')\sn(x+a') \big] 
+\cs(a)\ns(a')\cn(x)\dn(x)~,
\eeqa
\beqa\label{c69}
&&\m\sn(x)\cn(x)\dn(x+a)\cn(x+a')
=-\ds(a)\ds(a-a')
\big[\cs(a)\cn(x)-\ns(a)\cn(x+a) \big] \nonumber \\
&&\hspace{.4in}+\ds(a')\cs(a-a')
\big[\cs(a')\cn(x)-\ns(a')\cn(x+a') \big] 
-\cs(a)\ds(a')\sn(x)\dn(x)~.
\eeqa

\sss\noindent{\bf\large Rank 4 identities with 2 distinct arguments ($x, x+a$).}

\beqa
&&\m\dn(x)\sn(x)\cn(x)\dn(x+a)
=-\cs(a)[\ds^2(a)+1]\dn(x)
+\cs(a)\ds(a)\ns(a)\dn(x+a)\nonumber \\
&&\hspace{2.2in}+\,\cs(a)\dn^3(x)
+\m\ds(a)\ns(a)\cn(x)\sn(x)~,
\eeqa
\beqa
&&\m\dn(x)\sn(x)\cn(x)\sn(x+a)
=-\ns(a)[\ds^2(a)-1]\sn(x)+\cs(a)\ds(a)\ns(a)\sn(x+a) \nonumber \\
&&\hspace{2.2in}-\m\ns(a)\sn^3(x)
-\cs(a)\ds(a)\cn(x)\dn(x)~,
\eeqa
\beqa
&&\m\dn(x)\sn(x)\cn(x)\cn(x+a)
=-\ds(a)[\m+\cs^2(a)]\cn(x)
+\cs(a)\ds(a)\ns(a)\cn(x+a) \nonumber \\
&&\hspace{2.2in}+\m\ds(a)\cn^3(x)
+\cs(a)\ns(a)\sn(x)\dn(x)~,
\eeqa
\beqa
&&\dn^3(x)\dn(x+a)
=-\m\cs(a)\sn(x)\cn(x)\dn(x)
+\,\ds(a)\ns(a)\dn^2(x) \nonumber \\
&&\hspace{1.4in}-\cs^2(a) \dn(a)-\cs^3(a)\left [\Z(x+a)-\Z(x)-\Z(a) \right ]~,
\eeqa
\beqa
&&\dn^3(x)\sn(x+a)
=\cs^3(a)\cn(x+a)
-\ns(a)\left(2\cs^2(a)-\ds^2(a)\right)\cn(x) \nonumber \\
&&\hspace{1.4in}+\,\cs(a)\ds(a)\sn(x)\dn(x)
+\m\ns(a)\cn^3(x)~,
\eeqa
\beqa
&&\dn^3(x)\cn(x+a)
=-\cs^3(a)\sn(x+a)
-\ds(a)\left(2-\ns^2(a)\right)\sn(x) \nonumber \\
&&\hspace{1.4in}+\,\cs(a)\ns(a)\cn(x)\dn(x)
+\m\ds(a)\sn^3(x)~,
\eeqa
\beqa
&&m\sn^3(x)\dn(x+a)
=-\ns^3(a)\cn(x+a)+\left(\m+\ns^2(a)\right)\cs(a)\cn(x) \nonumber \\
&&\hspace{1.4in}-\,\ds(a)\ns(a)\sn(x)\dn(x)-\m\cs(a)\cn^3(x)~,
\eeqa
\beqa
&&\m^2\sn^3(x)\sn(x+a)
=-\,\m\ns(a)\sn(x)\cn(x)\dn(x)+\cs(a)\ds(a)\dn^2(x) \nonumber \\
&&\hspace{1.4in}-\cs(a)\ds(a) -\ns^3(a)\left [\Z(x+a)-\Z(x)-\Z(a) \right ]~,
\eeqa
\beqa
&&\m^2\sn^3(x)\cn(x+a)
=\left(1+\ns^2(a)\right)\ds(a)\dn(x)-\ns^3(a)\dn(x+a) \nonumber \\
&&\hspace{1.4in}-\,\m\cs(a)\ns(a)\sn(x)\cn(x)-\ds(a)\dn^3(x)~,
\eeqa
\beqa
&&\m\cn^3(x)\dn(x+a)
=-\ds^3(a)\sn(x+a)+\left(2\ds^2(a)-\ns^2(a)\right )\cs(a)
\sn(x) \nonumber \\
&&\hspace{1.4in}+\,\ds(a)\ns(a)\cn(x)\dn(x)+\m\cs(a)\sn^3(x)~,
\eeqa
\beqa
&&\m^2\cn^3(x)\sn(x+a)
=\ds^3(a)\dn(x+a)-\left(2\ds^2(a)-\cs^2(a)\right)\ns(a)\dn(x)\nonumber \\
&&\hspace{1.4in}+\,\m\cs(a)\ds(a)\sn(x)\cn(x)+\ns(a)\dn^3(x)~,
\eeqa
\beqa
&&\m^2\cn^3(x)\cn(x+a)
=-\,\m\ds(a)\sn(x)\cn(x)\dn(x)
+\cs(a)\ns(a)\dn^2(x) \nonumber \\
&&\hspace{.4in}-\cs(a)\ns(a)[1-\m+\m\dn^2(a)]-\ds^3(a)
\left [\Z(x+a)-\Z(x)-\Z(a) \right ]~,
\eeqa
\beqa\label{c15}
&&\dn^2(x)\dn^2(x+a)
=-\cs^2(a)[\dn^2(x)+\dn^2(x+a)]
+[\ds^2(a)+\cs^2(a)] \nonumber \\
&&\hspace{1.4in}+2\cs(a)\ds(a)\ns(a) \left [\Z(x+a)-\Z(x)-\Z(a) \right ]~,
\eeqa
\beqa
&&\m\dn^2(x)\sn(x+a) \cn(x+a)
=\cs(a)[\ds^2(a)+\ns^2(a)]\dn(x)-2\cs(a)\ds(a)\ns(a)\dn(x+a) \nonumber \\
&&\hspace{1.9in}-\,\m\cs^2(a)\sn(x+a)\cn(x+a)-\m\ds(a)\ns(a)\cn(x)\sn(x)~,
\eeqa
\beqa
&&\m\sn^2(x)\dn(x+a) \cn(x+a)
=\ns(a)[\cs^2(a)+\ds^2(a)]\sn(x)-2\cs(a)\ds(a)\ns(a)\sn(x+a) \nonumber \\
&&\hspace{1.9in}+\,\ns^2(a)\cn(x+a)\dn(x+a)+\cs(a)\ds(a)\cn(x)\dn(x)~,
\eeqa
\beqa
&&\m\cn^2(x)\dn(x+a) \sn(x+a)
=\ds(a)[\cs^2(a)+\ns^2(a)]\cn(x)-2\cs(a)\ds(a)\ns(a)\cn(x+a) \nonumber \\
&&\hspace{1.9in}-\,\ds^2(a)\sn(x+a)\dn(x+a)-\cs(a)\ns(a)\sn(x)\dn(x)~,
\eeqa
\beqa
&&\m^2\sn(x)\cn(x)\sn(x+a)\cn(x+a)
=\ds(a)\ns(a) \left [\dn^2(x)+\dn^2(x+a) \right ] \nonumber \\
&&\hspace{1.4in}-\left [\ds^2(a)+\ns^2(a) \right ]
\{\dn(a)+\cs(a)\left [\Z(x+a)-\Z(x)-\Z(a) \right ] \}~,
\eeqa
\beqa
&&\m\dn(x)\sn(x)\dn(x+a)\sn(x+a)
=-\cs(a) \ns(a)[1+\dn^2(a)]+\cs(a)\ns(a) \left [\dn^2(x)+\dn^2(x+a) \right ] \nonumber \\
&&\hspace{2.4in}
-\ds(a)[\cs^2(a)+\ns^2(a)] \left [\Z(x+a)-\Z(x)-\Z(a) \right ]~,
\eeqa
\beqa
&&\m\dn(x)\cn(x)\dn(x+a)\cn(x+a)
=2\cs(a)\ds(a)-\cs(a)\ds(a) \left [\dn^2(x)+\dn^2(x+a) \right ] \nonumber \\
&&\hspace{2.4in}
+\ns(a)[\ds^2(a)+\cs^2(a)] \left [\Z(x+a)-\Z(x)-\Z(a) \right ]~,
\eeqa
\beqa
&&\m\dn(x)\cn(x)\dn(x+a) \sn(x+a)
=\ds(a)[\cs^2(a)+\ns^2(a)]\dn(x)-\ns(a)[\cs^2(a)+\ds^2(a)]\dn(x+a) \nonumber \\
&&\hspace{2.in}-\,\m\cs(a)\ds(a)\sn(x+a)\cn(x+a)-\m\cs(a)\ns(a)\cn(x)\sn(x)~,
\eeqa
\beqa
&&\m\dn(x)\cn(x)\sn(x+a) \cn(x+a)
=\cs(a)[\ds^2(a)+\ns^2(a)]\cn(x)-\ns(a)[\cs^2(a)+\ds^2(a)]\cn(x+a) \nonumber \\
&&\hspace{2.in}-\,\cs(a)\ds(a)\sn(x+a)\dn(x+a)-\ds(a)\ns(a)\sn(x)\dn(x)~,
\eeqa
\beqa
&&\m\sn(x)\cn(x)\dn(x+a) \sn(x+a)
=\ds(a)[\cs^2(a)+\ns^2(a)]\sn(x)-\cs(a)[\ds^2(a)+\ns^2(a)]\sn(x+a) \nonumber \\
&&\hspace{2.in}+\,\ds(a)\ns(a)\cn(x+a)\dn(x+a)+\cs(a)\ns(a)\cn(x)\dn(x)~.
\eeqa

\sss
\noindent{\bf\Large Appendix D: Some Examples of Local Identities of Rank 5}

\beqa\label{dscds}
&&\m\dn(x)\sn(x)\cn(x)\dn(x+a) \sn(x+a)
=-\cs(a)\ns(a)[\cs^2(a)+\ds^2(a)+\ns^2(a)]\cn(x) \nonumber \\
&&\hspace{.2in}+\left[\ns^2(a)(\ds^2(a)+\cs^2(a))+\cs^2(a)\ds^2(a) \right ]\cn(x+a)
+\,\cs(a)\ds(a)\ns(a)\sn(x+a)\dn(x+a) \nonumber \\
&&\hspace{.2in}+\ds(a)[\cs^2(a)+\ns^2(a)]\sn(x)\dn(x) +\m \cs(a)\ns(a)\cn^3(x)~,
\eeqa
\beqa\label{dscdc}
&&\m\dn(x)\sn(x)\cn(x)\dn(x+a) \cn(x+a)
=\cs(a)\ds(a)[\cs^2(a)+\ds^2(a)+\ns^2(a)]\sn(x) \nonumber \\
&&\hspace{.2in}-\left[\ns^2(a)(\ds^2(a)+\cs^2(a))+\cs^2(a)\ds^2(a) \right ]\sn(x+a)
+\,\cs(a)\ds(a)\ns(a)\cn(x+a)\dn(x+a) \nonumber \\
&&\hspace{.2in}+\ns(a) [\cs^2(a)+\ds^2(a)]\cn(x)\dn(x) +\m \cs(a)\ds(a)\sn^3(x)~,
\eeqa
\beqa\label{dscsc}
&&\m^2 \dn(x)\sn(x)\cn(x)\sn(x+a) \cn(x+a)
=-\ds(a)\ns(a)[\cs^2(a)+\ds^2(a)+\ns^2(a)]\dn(x) \nonumber \\
&&\hspace{.2in}+\left[\ns^2(a)(\ds^2(a)+\cs^2(a))+\cs^2(a)\ds^2(a) \right ]\dn(x+a)
+\,\m\cs(a)\ds(a)\ns(a)\cn(x+a)\sn(x+a) \nonumber \\
&&\hspace{.2in}+\m\cs(a)[\ds^2(a)+\ns^2(a)]\cn(x)\sn(x) + \ds(a)\ns(a)\dn^3(x)~,
\eeqa
\beqa\label{dscdd}
&&\m\dn(x)\sn(x) \cn(x)\dn^2(x+a)
=\cs(a)\ds(a)\ns(a) \big [\dn^2(x+a)+2\dn^2(x)-\dn^2(a)-2 \big ]
 \nonumber \\
&&-\,\m\cs^2(a)\dn(x)\sn(x)\cn(x)
+\big [\ns^2(a) \left (\ds^2(a)+\cs^2(a) \right )+\cs^2(a) \ds^2(a) 
\big ]
\left [\Z(x+a)-\Z(x) -\Z(a) \right ]~.
\eeqa

\sss
\noindent{\bf\Large Appendix E: Local Identities of Arbitrary Rank}

\sss
\noindent In the following identities, $B \equiv -\cs^2(a),~B_1 \equiv \ns^2(a),~B_2 \equiv -\ds^2(a)$.

\beq
\dn^{2n}(x)\dn(x+a)
=B^{n}\dn(x+a)
+ \left [\ds(a) \ns(a)\dn(x) -\m \cs(a) \cn(x) \sn(x) \right ]
\sum_{k=1}^{n} B^{k-1} [\dn(x)]^{2(n-k)}~,
\eeq
\beqa
&&\dn^{2n+1}(x)\dn(x+a)
=B^{n} \big (\dn(a)+\cs(a)[\Z(x+a)-\Z(x)-\Z(a)] \big ) \nonumber \\
&&\hspace{.4in}+ \left [\ds(a) \ns(a) \dn(x)-\m \cs(a) \cn(x) \sn(x) \right ]
\sum_{k=1}^{n} B^{k-1} [\dn(x)]^{2(n-k)+1}~,
\eeqa
\beq
\m^{n}\sn^{2n}(x)\sn(x+a)
=B_1^{n} \sn(x+a)
-\left [\cs(a) \ds(a) \sn(x)+\ns(a) \cn(x) \dn(x) \right ]
\sum_{k=1}^{n} \m^{n-k} B_1^{k-1} [\sn(x)]^{2(n-k)}~,
\eeq
\beqa
&&\m^{n+1}\sn^{2n+1}(x)\sn(x+a)
= -B_1^{n} \ns(a) \left [\Z(x+a) -\Z(x) - \Z(a) \right ] \nonumber \\
&&\hspace{.4in}-\left [\cs(a) \ds(a) \sn(x)+\ns(a) \cn(x) \dn(x) \right ]
\sum_{k=1}^{n} \m^{n-k} B_1^{k-1} [\sn(x)]^{2(n-k)+1}~,
\eeqa
\beqa
&&\m^{n}\cn^{2n}(x)\cn(x+a)
=B_2^{n}\cn(x+a) \nonumber \\
&&\hspace{.4in}+\left [\cs(a) \ns(a) \cn(x) -\ds(a) \sn(x) \dn(x) \right ]
\sum_{k=1}^{n} \m^{n-k} B_2^{k-1} [\cn(x)]^{2(n-k)}~,
\eeqa
\beqa
&&\m^{n+1}\cn^{2n+1}(x)\cn(x+a)
=B_2^{n} \{\m\cn(a)+\ds(a) \left[\Z(x+a)-\Z(x)-\Z(a) \right ] 
\}\nonumber \\
&&\hspace{.4in}+\left [\cs(a) \ns(a) \cn(x) -\ds(a) \sn(x) \dn(x) \right ]
\sum_{k=1}^{n} \m^{n-k} B_2^{k-1} [\cn(x)]^{2(n-k)+1}~,
\eeqa
\beqa
&&\m^{n}\cn^{2n}(x)\sn(x)\dn(x+a)
=- B_2^{n} \ns(a) \cn(x+a) +\cs(a) \m^{n} \cn^{2n+1} (x) \nonumber \\
&&\hspace{.4in}-\ns(a) \left [\cs(a) \ns(a) \cn(x) - \ds(a) \sn(x) \dn(x) \right ]
\sum_{k=1}^{n} \m^{n-k} B_2^{k-1} [\cn(x)]^{2(n-k)}~,
\eeqa
\beqa
&&\m^{n}\sn^{2n}(x) \cn(x) \dn(x+a)
=B_1^{n} \ds(a) \sn(x+a) -\cs(a) \m^{n} \sn^{2n+1} (x) \nonumber \\
&&\hspace{.4in}-\ds(a) \left [\cs(a) \ds(a) \sn(x)+\ns(a) \cn(x) \dn(x) \right ]
\sum_{k=1}^{n} \m^{n-k} B_1^{k-1} [\sn(x)]^{2(n-k)}~,
\eeqa
\beqa
&&\m\dn^{2n}(x) \cn(x) \sn(x+a)
=-B^{n} \ds(a) \dn(x+a) +\ns(a) \dn^{2n+1} (x) \nonumber \\
&&\hspace{.4in}- \ds(a) \left [\ds(a) \ns(a)\dn(x) -\m \cs(a) \cn(x) \sn(x) \right ]
\sum_{k=1}^{n} B^{k-1} [\dn(x)]^{2(n-k)}~,
\eeqa
\beqa
&&\m\dn^{2n}(x)\sn(x+a)\cn(x+a)
=\m B^{n} \sn(x+a)\cn(x+a)
-2nB^{n-1} \ds(a)\cs(a)\ns(a) \dn(x+a) \nonumber \\
&&\hspace{1.4in}+\sum_{k=1}^{n} B^{k-1}
\left [(m+2\ds^2(a))\cs(a)+2(k-1)\ds^2(a)\nc(a) \ns(a) \right ]
[\dn(x)]^{2(n-k)+1} \nonumber \\
&&\hspace{1.4in}- m\ds(a) \ns(a) \sum_{k=1}^{n} (2k-1) B^{k-1}
\cn(x)\sn(x)[\dn(x)]^{2(n-k)}~,
\eeqa
\beqa
&&\m\dn^{2n+1}(x)\sn(x+a)\cn(x+a)
=- m\ds(a) \ns(a) \sum_{k=1}^{n} (2k-1) B^{k-1}
\cn(x)\sn(x)[\dn(x)]^{2(n-k)+1} \nonumber \\
&&\hspace{.1in}+B^{n-1}\cs(a)
\bigg[(1-m)-(2n+1)\ds(a)\ns(a)
\big (\dn(a)+\cs(a)[\Z(x+a)-\Z(x)-\Z(a)] \big ) \bigg ] \nonumber \\
&&\hspace{.1in}+\sum_{k=1}^{n}\! B^{k-1}
[(\m+2\ds^2(a))\cs(a)+2(k-1)\ds^2(a)\nc(a) \ns(a)]
 [\dn(x)]^{2(n-k+1)}\!-\!B^{n}\cs(a)\dn^{2}(x+a)~,
\eeqa
\beqa\label{d26}
&&\dn^{2n}(x)\sn(x+a)\dn(x+a)
= B^{n} \sn(x+a)\dn(x+a)
-2nB^{n-1} \ds(a)\cs(a)\ns(a) \cn(x+a) \nonumber \\
&&\hspace{1.4in}-\sum_{k=1}^{n} B^{k-1} \left [\cs(a)\ns(a)+2(k-1)\ds^2(a)\nc(a) \right ]
 \sn(x) [\dn(x)]^{2(n-k)+1} \nonumber \\
&&\hspace{1.4in}+\ds(a) \sum_{k=1}^{n}  B^{k-1} [\cs^2(a)+(2k-1)\ns^2(a)]
\cn(x)[\dn(x)]^{2(n-k)}~,
\eeqa
\beqa
&&\dn^{2n+1}(x)\sn(x+a)\dn(x+a)
= -B^{n} \cs(a) \cn(x+a)\dn(x+a)
+B^{n-1} \cs(a)\ns(a)[\cs^2(a)+2n\ds^2(a)]\sn(x) \nonumber \\
&&\hspace{.2in}-\sum_{k=1}^{n} B^{k-1} [\cs(a)\ns(a)+2(k-1)\ds^2(a)\nc(a)]
\sn(x) [\dn(x)]^{2(n-k+1)}+(2n+1)B^{n} \ds(a)\ns(a) \sn(x+a) \nonumber \\
&&\hspace{1.2in}+\ds(a) \sum_{k=1}^{n}  B^{k-1} [\cs^2(a)+(2k-1)\ns^2(a)]
\cn(x)[\dn(x)]^{2(n-k)+1}~,
\eeqa
\beqa
&&\dn^{2n}(x)\cn(x+a)\dn(x+a)
= B^{n} \cn(x+a)\dn(x+a)
+2nB^{n-1} \ds(a)\cs(a)\ns(a) \sn(x+a) \nonumber \\
&&\hspace{1.4in}-\sum_{k=1}^{n} B^{k-1} \left [\cs(a)\ds(a)+2(k-1)\ds(a)\nc(a)\ns(a) \right ]
\cn(x) [\dn(x)]^{2(n-k)+1} \nonumber \\
&&\hspace{1.4in}-\ns(a) \sum_{k=1}^{n}  B^{k-1} [\cs^2(a)+(2k-1)\ds^2(a)]
\sn(x)[\dn(x)]^{2(n-k)}~,
\eeqa
\beqa
&&\dn^{2n+1}(x) \cn(x+a)\dn(x+a)
= B^{n} \cs(a) \sn(x+a)\dn(x+a)
+B^{n-1} \cs(a)\ds(a)[\cs^2(a)+2n\ns^2(a)]\cn(x) \nonumber \\
&&\hspace{.2in}-\sum_{k=1}^{n} B^{k-1} [\cs(a)\ds(a)+2(k-1)\ds(a)\nc(a)\ns(a) ]
\cn(x) [\dn(x)]^{2(n-k+1)}\!+(2n+1)B^{n} \ds(a)\ns(a) \cn(x+a) \nonumber \\
&&\hspace{1.2in}-\ns(a) \sum_{k=1}^{n}\!  B^{k-1} [\cs^2(a)+(2k-1)\ds^2(a)]
\sn(x)[\dn(x)]^{2(n-k)+1}~,
\eeqa
\beqa\label{d15}
&&\dn^{2n}(x)\dn^2(x+a)
=2nB^{n-1} \ds(a)\ns(a)
[\dn(a)+\cs(a)(\Z(x+a)-\Z(x)-\Z(a) )] \nonumber \\
&&-(1-m)B^{n-1}
+ \sum_{k=1}^{n-1} B^{k-1} [B^2-(1-m)+2k\ds^2(a)\ns^2(a)]
[\dn(x)]^{2(n-k)} \nonumber \\
&&+B^{n} \dn^{2} (x+a)+B\dn^{2n}-2m\cs(a)\ds(a)\ns(a) \sn(x) \cn(x)
\sum_{k=1}^{n-1} k B^{k-1}
[\dn(x)]^{2(n-k)-1}~,
\eeqa
\beqa
&&\dn^{2n+1}(x)\dn^2(x+a)
=B\dn^{2n+1} (x)+B^{n} \m \cs(a) \cn(x+a) \sn(x+a) \nonumber \\
&&+ \sum_{k=1}^{n} B^{k-1} [B^2-(1-m)+2k\ds^2(a)\ns^2(a)]
[\dn(x)]^{2(n-k)+1} \nonumber \\
&&-2m\cs(a)\ds(a)\ns(a) \sn(x) \cn(x)
\sum_{k=1}^{n} k B^{k-1}
[\dn(x)]^{2(n-k)}
+(2n+1) B^{n} \ds(a) \ns(a) \dn(x+a)~,
\eeqa

\sss\noindent{\bf\Large Appendix F: Examples of Identities with Weighted Terms and Their Linear Combinations}
\sss 

\noindent In this appendix $a=\f{2rK}{p}, a'=\f{2sK}{p}, b=\f{4rK}{p}$. 

{\bf F1: MI-I Identities}

\beq\label{e1}
\m\sum_{j=1}^{p} c_j [s_{j+r}-s_{j-r}] =
2 [\ns(a) -\ds(a)]\sum_{j=1}^{p} d_j~,
\eeq
\beq\label{e2}
\sum_{j=1}^{p} d^2_j [d_{j+r}-d_{j-r}] =
-2\m \cs(a)\sum_{j=1}^{p} c_j s_j~,
\eeq
\beq\label{ee2}
\sum_{j=1}^{p} c_j [c_{j+r}d_{j+r}-c_{j-r}d_{j-r}] =
2 \ds(a)\sum_{j=1}^{p} c_j s_j~,
\eeq
\beq\label{ee3}
\sum_{j=1}^{p} s_j [s_{j+r}d_{j+r}-s_{j-r}d_{j-r}] =
-2 \ns(a)\sum_{j=1}^{p} c_j s_j~,
\eeq

\beq\label{e3}
\sum_{j=1}^{p} c_j [d_{j+r}c_{j+s}-d_{j-r}c_{j-s}] = 0~,
\eeq
\beq\label{e4}
\sum_{j=1}^{p} s_j [d_{j+r}s_{j+s}-d_{j-r}s_{j-s}] = 0~,
\eeq
\beq\label{e5}
\sum_{j=1}^{p} d_j [d_{j+r}d_{j+s}-d_{j-r}d_{j-s}] = 0~,
\eeq
\beq\label{e6}
\sum_{j=1}^{p} d_j [c_{j+r}c_{j+s}-c_{j-r}c_{j-s}] = 0~,
\eeq
\beq\label{e7}
\sum_{j=1}^{p} d_j [s_{j+r}s_{j+s}-s_{j-r}s_{j-s}] = 0~,
\eeq
\beq\label{e8}
\m\sum_{j=1}^{p} d^2_j [c_{j+r} s_{j+r}-c_{j-r} s_{j-r}] =
2 \cs(a)[\ns^2(a)+\ds^2(a)-2\ns(a)\ds(a)] \sum_{j=1}^{p} d_j~,
\eeq
\beq\label{e9}
\m\sum_{j=1}^{p} c_j d_j [d_{j+r} s_{j+r}-d_{j-r} s_{j-r}] =
-2 \big [\ns(a)(\cs^2(a)+\ds^2(a))-\ds(a)(\cs^2(a)+\ns^2(a)) \big] 
\sum_{j=1}^{p} d_j~,
\eeq
\beq\label{e10}
\sum_{j=1}^{p} d^3_j [d^2_{j+r}-d^2_{j-r}] =
-2\m \cs(a)[\cs^2(a)+2\ds(a)\ns(a)]\sum_{j=1}^{p} c_j s_j~,
\eeq
\beq\label{e11}
m\sum_{j=1}^{p} c_j s_j d_j [c_{j+r}s_{j+r}-c_{j-r}s_{j-r}] =
2 \cs(a)\big [(\ds^2(a)+\ns^2(a))+\ds(a)\ns(a) \big ]\sum_{j=1}^{p} c_j s_j~,
\eeq
\beqa\label{e12}
&&m\sum_{j=1}^{p} c_j s_j d_j [d_{j+r}-d_{j-r}] 
=2\cs(a)\sum_{j=1}^{p} d_j^3 \nonumber \\
&&-2 \cs(a)\big [\ds^2(a)+1-\ds(a)\ns(a) \big ]\sum_{j=1}^{p} d_j~,
\eeqa
\beqa\label{e13}
&&m\sum_{j=1}^{p} c_j s_j d_j [d^3_{j+r}-d^3_{j-r}] 
=2\cs(a)[\ds(a)\ns(a)-\cs^2(a)]\sum_{j=1}^{p} d_j^3 
+2 \cs(a)\bigg [2\cs^2(a)\ds^2(a) \nonumber \\
&&+2\ns^2(a)(\ds^2(a)+\cs^2(a))-\cs^4(a)
-\ds(a)\ns(a)(\ds^2(a)+\cs^2(a)+\ns^2(a)) \bigg ]\sum_{j=1}^{p} d_j~,
\eeqa
\beq\label{e14}
\m\sum_{j=1}^{p} c_j d_j [d_{j+r}s_{j+s}-d_{j-r}s_{j-s}] =
2[\ns(a-a')\cs(a)(\ds(a)-\ns(a)) 
-\cs(a-a')\cs(a')(\ds(a')-\ns(a'))] \sum_{j=1}^{p} d_j~, 
\eeq
\beq\label{e15}
\m\sum_{j=1}^{p} s_j d_j [d_{j+r}c_{j+s}-d_{j-r}c_{j-s}] =
-2[\ds(a-a')\cs(a)(\ds(a)-\ns(a)) 
-\cs(a-a')\cs(a')(\ds(a')-\ns(a'))] \sum_{j=1}^{p} d_j~, 
\eeq
\beq\label{e16}
\m\sum_{j=1}^{p} c_j s_j [d_{j+r}d_{j+s}-d_{j-r}d_{j-s}] =
2\cs(a-a')[\ns(a) (\ds(a)-\ns(a))-\ns(a') (\ds(a')-\ns(a'))] 
\sum_{j=1}^{p} d_j~, 
\eeq
\beq\label{e17}
\m^2\sum_{j=1}^{p} c_j s_j [s_{j+r}s_{j+s}-s_{j-r}s_{j-s}] =
-2\ns(a-a')[\ns(a) (\ds(a)-\ns(a))-\ns(a') (\ds(a')-\ns(a'))] 
\sum_{j=1}^{p} d_j~, 
\eeq
\beq\label{e18}
\m^2\sum_{j=1}^{p} c_j s_j [c_{j+r}c_{j+s}-c_{j-r}c_{j-s}] =
2\ds(a-a')[\ns(a) (\ds(a)-\ns(a))-\ns(a') (\ds(a')-\ns(a'))]
\sum_{j=1}^{p} d_j~, 
\eeq
\beq\label{e10a}
\sum_{j=1}^{p} d^4_j [d_{j+r}-d_{j-r}] =
-2\m \cs(a) \sum_{j=1}^{p} c_j s_j d_j^2
+2\m \cs^{3}(a) \sum_{j=1}^{p} c_j s_j~,
\eeq
\beq\label{e10b}
\m^2\sum_{j=1}^{p} c^3_j [s_{j+r}-s_{j-r}] =
2 \ns(a) \sum_{j=1}^{p} d_j^3
-2 [\ns(a)(2\ds^2(a)-\cs^2(a))-\ds^3(a)] \sum_{j=1}^{p} d_j~,
\eeq
\beq\label{e10c}
\m^2\sum_{j=1}^{p} s^3_j [c_{j+r}-c_{j-r}] =
-2 \ds(a) \sum_{j=1}^{p} d_j^3
+2 [\ds(a)(\ns^2(a)+1)-\ns^3(a)] \sum_{j=1}^{p} d_j~,
\eeq

{\bf F2: MI-II Identities}

\beq\label{e19}
\m\sum_{j=1}^{p} s_j d_j [c_{j+r}-c_{j-r}] =
2 \ds(a)\sum_{j=1}^{p} d^2_j -2p\ns(a)[\dn(a) -\cs(a) \Z(a)]~,
\eeq
\beq\label{e20}
\m\sum_{j=1}^{p} c_j d_j [s_{j+r}-s_{j-r}] =
2 \ns(a)\sum_{j=1}^{p} d^2_j -2p\ds(a)[\dn(a) -\cs(a) \Z(a)]~,
\eeq
\beq\label{e21}
\m\sum_{j=1}^{p} s_j c_j [d_{j+r}-d_{j-r}] =
2 \cs(a)\sum_{j=1}^{p} d^2_j -2p[\cs(a) -\ds(a) \ns(a) \Z(a)]~,
\eeq
\beq\label{e22}
\m\sum_{j=1}^{p} d_j [s_{j+r}c_{j+s}-s_{j-r}c_{j-s}] =
-2p\ds(a-a')[\dn(a) -\cs(a) \Z(a)]
+2p\ns(a-a')[\dn(a') -\cs(a') \Z(a')]~,
\eeq
\beq\label{e23}
\m\sum_{j=1}^{p} c_j [d_{j+r}s_{j+s}-d_{j-r}s_{j-s}] =
2p\dc(a)\ns(a-a')[\dn(a) -\cs(a) \Z(a)]
-2p\dc(a')\cs(a-a')[\dn(a') -\cs(a') \Z(a')]~,
\eeq
\beq\label{e24}
\m\sum_{j=1}^{p} s_j [d_{j+r}c_{j+s}-d_{j-r}c_{j-s}] =
2p\nc(a)\ds(a-a')[\dn(a) -\cs(a) \Z(a)]
-2p\nc(a')\cs(a-a')[\dn(a') -\cs(a') \Z(a')]~,
\eeq
\beq\label{e25}
\sum_{j=1}^{p} d^3_j [d_{j+r}-d_{j-r}] =
-2\m  \cs(a)\sum_{j=1}^{p} d_j s_j c_j~,
\eeq
\beq\label{e26}
\m\sum_{j=1}^{p} s^3_j [s_{j+r}-s_{j-r}] =
-2  \ns(a)\sum_{j=1}^{p} d_j s_j c_j~,
\eeq
\beq\label{e27}
\m \sum_{j=1}^{p} c^3_j [c_{j+r}-c_{j-r}] =
-2  \ds(a)\sum_{j=1}^{p} d_j s_j c_j~,
\eeq
\beqa\label{e28}
&&\m\sum_{j=1}^{p} d^2_j [d_{j+r}s_{j+r}c_{j+r}-d_{j-r}s_{j-r}c_{j-r}] =
-6\cs(a)\ds(a)\ns(a) \sum_{j=1}^{p} d_j^2 \nonumber \\
&&+2p\cs(a)\ds(a)\ns(a)[\dn^2(a)+2]-2p[\ns^2(a)(\cs^2(a)+\ds^2(a))
+\cs^2(a)\ds^2(a)]\Z(a)~,
\eeqa
\beqa\label{e29}
&&\m\sum_{j=1}^{p} s_j c_j [d^3_{j+r}-d^3_{j-r}] =
-2\cs(a)[\ds^2(a)+\ns^2(a)+\cs^2(a)] \sum_{j=1}^{p} d^2_j \nonumber \\
&&-2p\cs(a)[1-m-3\ds^2(a) +3\cs(a)\ds(a)\ns(a)\Z(a)]~,
\eeqa
\beqa\label{e30}
&&\m^2\sum_{j=1}^{p} d_j c_j [s^3_{j+r}-s^3_{j-r}] =
2\ns(a)[\ds^2(a)+\ns^2(a)+\cs^2(a)] \sum_{j=1}^{p} d^2_j \nonumber \\
&&+2p\ns(a)[1-m-3\ds^2(a) +3\cs(a)\ds(a)\ns(a)\Z(a)]~,
\eeqa
\beqa\label{e31}
&&\m^2\sum_{j=1}^{p} d_j s_j [c^3_{j+r}-c^3_{j-r}] =
-2\ds(a)[\ds^2(a)+\ns^2(a)+\cs^2(a)] \sum_{j=1}^{p} d^2_j \nonumber \\
&&-2p\ds(a)[1-m-3\ds^2(a) +3\cs(a)\ds(a)\ns(a)\Z(a)]~,
\eeqa

{\bf F3: MI-III Identities}

\beq\label{e32}
\sum_{j=1}^{p} d_j [c_{j+r}-c_{j-r}] =
2 [\cs(b) -\ds(b)]\sum_{j=1}^{p} s_j~,
\eeq
\beq\label{e33}
\sum_{j=1}^{p} \m s^2_j [s_{j+r}-s_{j-r}] =
-2 \ns(b)\sum_{j=1}^{p} c_j d_j~,
\eeq
\beq\label{e34}
\sum_{j=1}^{p} c_j [s_{j+r}c_{j+s}-s_{j-r}c_{j-s}] = 0~,
\eeq
\beq\label{e35}
\sum_{j=1}^{p} d_j [d_{j+r}s_{j+s}-d_{j-r}s_{j-s}] = 0~,
\eeq
\beq\label{e36}
\sum_{j=1}^{p} s_j [d_{j+r}d_{j+s}-d_{j-r}d_{j-s}] = 0~,
\eeq
\beq\label{e37}
\sum_{j=1}^{p} s_j [c_{j+r}c_{j+s}-c_{j-r}c_{j-s}] = 0~,
\eeq
\beq\label{e38}
\sum_{j=1}^{p} s_j [s_{j+r}s_{j+s}-s_{j-r}s_{j-s}] = 0~,
\eeq
\beq\label{e39}
\sum_{j=1}^{p} d^2_j [c_{j+r} d_{j+r}-c_{j-r} d_{j-r}] =
-2 \ns(b)[\cs^2(b)+\ds^2(b)-2\cs(b)\ds(b)] \sum_{j=1}^{p} s_j~,
\eeq
\beq\label{e40}
\m\sum_{j=1}^{p} s_j d_j [c_{j+r} s_{j+r}-c_{j-r} s_{j-r}] =
-2 \big [\ds(b)(\cs^2(b)+\ns^2(b))-\cs(b)(\ds^2(b)+\ns^2(b)) \big] 
\sum_{j=1}^{p} s_j~,
\eeq
\beq\label{e41}
\m^2\sum_{j=1}^{p} s^3_j [s^2_{j+r}-s^2_{j-r}] =
2 \ns(b)[\ns^2(b)+2\ds(b)\cs(b)]\sum_{j=1}^{p} c_j d_j~,
\eeq
\beq\label{e42}
m\sum_{j=1}^{p} c_j s_j d_j [c_{j+r}d_{j+r}-c_{j-r}d_{j-r}] =
2 \ns(b)\big [(\ds^2(b)+\cs^2(b))+\ds(b)\cs(b) \big ]\sum_{j=1}^{p} c_j d_j~,
\eeq
\beqa\label{e43}
&&m\sum_{j=1}^{p} c_j s_j d_j [s_{j+r}-s_{j-r}] 
=-2\m \ns(b)\sum_{j=1}^{p} s_j^3 \nonumber \\
&&-2 \ns(b)\big [(-\ds^2(b)+1))+\ds(b)\ns(b) \big ]\sum_{j=1}^{p} s_j~,
\eeqa
\beqa\label{e44}
&&\m^2\sum_{j=1}^{p} c_j s_j d_j [s^3_{j+r}-s^3_{j-r}] 
=2\m \ns(b)[\ds(b)\cs(b)-\ns^2(b)]\sum_{j=1}^{p} s_j^3 
-2 \ns(b)\bigg [3\cs^2(b)\ds^2(b) \nonumber \\
&&+\ns^2(b)(2\ds^2(b)+\cs^2(b))
-2\ds(b)\cs(b)(\ds^2(b)+\cs^2(b)+\ns^2(b)) \bigg ]\sum_{j=1}^{p} s_j~,
\eeqa
\beq\label{e45}
\m\sum_{j=1}^{p} c_j s_j [d_{j+r}s_{j+s}-d_{j-r}s_{j-s}] =
2\big [\ns(b-b')\ns(b)(\ds(b)-\cs(b)) 
-\cs(b-b')\ns(b')(\ds(b')-\cs(b')) \big ] \sum_{j=1}^{p} s_j~, 
\eeq
\beq\label{e46}
\m\sum_{j=1}^{p} s_j d_j [s_{j+r}c_{j+s}-s_{j-r}c_{j-s}] =
-2 \big [\ds(b-b')\ns(b)(\ds(b)-\cs(b)) 
-\ns(b-b')\ns(b')(\ds(b')-\cs(b')) \big ] \sum_{j=1}^{p} s_j~, 
\eeq
\beq\label{e47}
\sum_{j=1}^{p} c_j d_j [d_{j+r}d_{j+s}-d_{j-r}d_{j-s}] =
-2\cs(b-b') \big [\cs(b) (\ds(b)-\cs(b))-\cs(b') (\ds(b')-\cs(b')) \big ] 
\sum_{j=1}^{p} s_j~,
\eeq
\beq\label{e48}
\m\sum_{j=1}^{p} c_j d_j [s_{j+r}s_{j+s}-s_{j-r}s_{j-s}] =
-2\ns(b-b') \big [\ds(b) (\ds(b)-\cs(b))-\ds(b') (\ds(b')-\cs(b')) \big ] 
\sum_{j=1}^{p} s_j~,
\eeq
\beq\label{e49}
\m\sum_{j=1}^{p} c_j d_j [c_{j+r}c_{j+s}-c_{j-r}c_{j-s}] =
2\ds(b-b') \big [\ds(b) (\ds(b)-\cs(b))-\ds(b') (\ds(b')-\cs(b')) \big ] 
\sum_{j=1}^{p} s_j~,
\eeq
\beq\label{e41a}
\m^2\sum_{j=1}^{p} s^4_j [s_{j+r}-s_{j-r}] =
2 \ns(b) \sum_{j=1}^{p} c_j d^3_j
-2 \ns(b)[\ns^2(b)+1]\sum_{j=1}^{p} c_j d_j~,
\eeq
\beq\label{e41b}
\sum_{j=1}^{p} d^3_j [c_{j+r}-c_{j-r}] =
2\m \ds(b) \sum_{j=1}^{p} s^3_j
-2 [\cs^3(b)-\ds(b)(\ns^2(b)-2]\sum_{j=1}^{p} s_j~,
\eeq
\beq\label{e41c}
\m\sum_{j=1}^{p} c^3_j [d_{j+r}-d_{j-r}] =
2\m \cs(b) \sum_{j=1}^{p} s^3_j
-2 [\ds^3(b)-\cs(b)(2\ds^2(b)-1]\sum_{j=1}^{p} s_j~,
\eeq

{\bf F4: MI-IV Identities}

\beq\label{e50}
\sum_{j=1}^{p} d_j [s_{j+r}-s_{j-r}] =
2 [\ns(b) -\cs(b)]\sum_{j=1}^{p} c_j~,
\eeq
\beq\label{e51}
\sum_{j=1}^{p} \m c^2_j [c_{j+r}-c_{j-r}] =
-2 \ds(b)\sum_{j=1}^{p} s_j d_j~,
\eeq
\beq\label{e52}
\sum_{j=1}^{p} s_j [s_{j+r}c_{j+s}-s_{j-r}c_{j-s}] = 0~,
\eeq
\beq\label{e53}
\sum_{j=1}^{p} d_j [d_{j+r}c_{j+s}-d_{j-r}c_{j-s}] = 0~,
\eeq
\beq\label{e54}
\sum_{j=1}^{p} c_j [d_{j+r}d_{j+s}-d_{j-r}d_{j-s}] = 0~,
\eeq
\beq\label{e55}
\sum_{j=1}^{p} c_j [c_{j+r}c_{j+s}-c_{j-r}c_{j-s}] = 0~,
\eeq
\beq\label{e56}
\sum_{j=1}^{p} c_j [s_{j+r}s_{j+s}-s_{j-r}s_{j-s}] = 0~,
\eeq
\beq\label{e57}
\sum_{j=1}^{p} d^2_j [s_{j+r} d_{j+r}-s_{j-r} d_{j-r}] =
2 \ds(b)[\cs^2(b)+\ns^2(b)-2\cs(b)\ns(b)] \sum_{j=1}^{p} c_j~,
\eeq
\beq\label{e58}
\m\sum_{j=1}^{p} c_j d_j [c_{j+r} s_{j+r}-c_{j-r} s_{j-r}] =
2 \big [\cs(b)(\ds^2(b)+\ns^2(b))-\ns(b)(\ds^2(b)+\cs^2(b)) \big] 
\sum_{j=1}^{p} c_j~,
\eeq
\beq\label{e59}
\m^2\sum_{j=1}^{p} c^3_j [c^2_{j+r}-c^2_{j-r}] =
-2 \ds(b)[\ds^2(b)+2\ns(b)\cs(b)]\sum_{j=1}^{p} s_j d_j~,
\eeq
\beq\label{e60}
m\sum_{j=1}^{p} c_j s_j d_j [s_{j+r}d_{j+r}-s_{j-r}d_{j-r}] =
2 \ds(b)\big [(\ns^2(b)+\cs^2(b))+\ns(b)\cs(b) \big ]\sum_{j=1}^{p} s_j d_j~,
\eeq
\beqa\label{e61}
&&m\sum_{j=1}^{p} c_j s_j d_j [c_{j+r}-c_{j-r}] 
=2\m \ds(b)\sum_{j=1}^{p} c_j^3 \nonumber \\
&&-2 \ds(b)\big [(\cs^2(b)+m))-\cs(b)\ns(b) \big ]\sum_{j=1}^{p} c_j~,
\eeqa
\beqa\label{e62}
&&\m^2\sum_{j=1}^{p} c_j s_j d_j [c^3_{j+r}-c^3_{j-r}] 
=2\m \ds(b)[\ns(b)\cs(b)-\ds^2(b)]\sum_{j=1}^{p} c_j^3 
-2 \ds(b)\bigg [2\cs(b)\ns(b)(\ds^2(b)+\cs^2(b) \nonumber \\
&&+\ns^2(b))-((2\ds^2(b)+3\cs^2(b))\ns^2(b)+\ds^2(b)(2\cs^2(b)-\ds^2(b)) \bigg ]
\sum_{j=1}^{p} c_j~,
\eeqa
\beq\label{e63}
\m\sum_{j=1}^{p} c_j s_j [d_{j+r}c_{j+s}-d_{j-r}c_{j-s}] =
2\big [\ds(b-b')\ds(b)(\ns(b)-\cs(b)) 
-\cs(b-b')\ds(b')(\ns(b')-\cs(b')) \big ] \sum_{j=1}^{p} c_j~, 
\eeq
\beq\label{e64}
\m\sum_{j=1}^{p} c_j d_j [s_{j+r}c_{j+s}-s_{j-r}c_{j-s}] =
-2 \big [\ds(b-b')\ds(b)(\ns(b)-\cs(b)) 
-\ns(b-b')\ds(b')(\ns(b')-\cs(b')) \big ] \sum_{j=1}^{p} c_j~, 
\eeq
\beq\label{e65}
\sum_{j=1}^{p} d_j s_j [d_{j+r}d_{j+s}-d_{j-r}d_{j-s}] =
2\cs(b-b') \big [\cs(b) (\ns(b)-\cs(b))-\cs(b') (\ns(b')-\cs(b')) \big ] 
\sum_{j=1}^{p} c_j~,
\eeq
\beq\label{e66}
\m\sum_{j=1}^{p} d_j s_j [s_{j+r}s_{j+s}-s_{j-r}s_{j-s}] =
2\ns(b-b') \big [\ns(b) (\ns(b)-\cs(b))-\ns(b') (\ns(b')-\cs(b')) \big ] 
\sum_{j=1}^{p} c_j~,
\eeq
\beq\label{e67}
\m\sum_{j=1}^{p} d_j s_j [c_{j+r}c_{j+s}-c_{j-r}c_{j-s}] =
-2\ds(b-b') \big [\ns(b) (\ns(b)-\cs(b))-\ns(b') (\ns(b')-\cs(b')) \big ] 
\sum_{j=1}^{p} c_j~,
\eeq
\beq\label{e68a}
\m^2\sum_{j=1}^{p} c^4_j [c_{j+r}-c_{j-r}] =
-2 \ds(b) \sum_{j=1}^{p} s_j d^3_j
+2 \ds(b)[2\ds^2(b)-\cs^2(b)]\sum_{j=1}^{p} s_j d_j~,
\eeq
\beq\label{e68b}
\sum_{j=1}^{p} d^3_j [s_{j+r}-s_{j-r}] =
2\m \ns(b) \sum_{j=1}^{p} c^3_j
+2 [\cs^3(b)+\ns(b)(\ds^2(b)-2\cs^2(b)]\sum_{j=1}^{p} c_j~,
\eeq
\beq\label{e68c}
\m\sum_{j=1}^{p} s^3_j [d_{j+r}-d_{j-r}] =
-2\m \cs(b) \sum_{j=1}^{p} c^3_j
-2 [\ns^3(b)-\cs(b)(\ns^2(b)+m]\sum_{j=1}^{p} c_j~,
\eeq

{\bf F5: MI-I Identities With Alternating Signs}

Note in this and the next subsection, p is even and r is necessarily an odd 
integr.

\beq\label{e70a}
\m\sum_{j=1}^{p} (-1)^{j-1} s_j [c_{j+r}-c_{j-r}] =
2 (\ds(a)+\ns(a)) \sum_{j=1}^{p} (-1)^{j-1} d_j~,
\eeq
\beq\label{e69}
\sum_{j=1}^{p} (-1)^{j-1} d^2_j [d_{j+r}-d_{j-r}] =
-2\m \cs(a) \sum_{j=1}^{p} (-1)^{j-1} c_j s_j~,
\eeq
\beq\label{e70}
\sum_{j=1}^{p} (-1)^{j-1} c_j d_j [c_{j+r}-c_{j-r}] =
-2 \ds(a) \sum_{j=1}^{p} (-1)^{j-1} c_j s_j~,
\eeq
\beq\label{e71}
\sum_{j=1}^{p} (-1)^{j-1} s_j d_j [s_{j+r}-s_{j-r}] =
2 \ns(a) \sum_{j=1}^{p} (-1)^{j-1} c_j s_j~,
\eeq
\beq\label{e72}
\sum_{j=1}^{p} (-1)^{j-1} d^2_j [c_{j+r}s_{j+r}-c_{j-r}s_{j-r}] =
2 \cs(a)[\ds^2(a)+\ns^2(a)+2\ds(a)\ns(a)] \sum_{j=1}^{p} (-1)^{j-1} d_j~,
\eeq
\beq\label{e73}
\sum_{j=1}^{p} (-1)^{j-1} s_j d_j [c_{j+r}d_{j+r}-c_{j-r}d_{j-r}] =
2\big [\ns(a)(\cs^2(a)+\ds^2(a)-\ds(a)(\cs^2(a)+\ns^2(a) \big ]
 \sum_{j=1}^{p} (-1)^{j-1} c_j s_j~,
\eeq

{\bf F6: MI-II Identities with Alternating Signs}

\beq\label{e74}
\m\sum_{j=1}^{p} (-1)^{j-1} c_j s_j [d_{j+r}-d_{j-r}] =
2\cs(a) \sum_{j=1}^{p} (-1)^{j-1} d^2_j~,
\eeq
\beq\label{e75}
\m\sum_{j=1}^{p} (-1)^{j-1} d_j s_j [c_{j+r}-c_{j-r}] =
2\ds(a) \sum_{j=1}^{p} (-1)^{j-1} d^2_j~,
\eeq
\beq\label{e76}
\m\sum_{j=1}^{p} (-1)^{j-1} c_j d_j [s_{j+r}-s_{j-r}] =
2\ns(a) \sum_{j=1}^{p} (-1)^{j-1} d^2_j~,
\eeq
\beq\label{e77}
\m\sum_{j=1}^{p} (-1)^{j-1} c_j s_j [d^3_{j+r}-d^3_{j-r}] =
2\cs(a)(\ds^2(a)+1) \sum_{j=1}^{p} (-1)^{j-1} d^2_j~,
\eeq
\beq\label{e78}
\m^2\sum_{j=1}^{p} (-1)^{j-1} d_j s_j [c^3_{j+r}-c^3_{j-r}] =
2\ds(a) (\cs^2(a)+m)\sum_{j=1}^{p} (-1)^{j-1} d^2_j~,
\eeq
\beq\label{e79}
\m\sum_{j=1}^{p} (-1)^{j-1} c_j d_j [s^3_{j+r}-s^3_{j-r}] =
2\ns(a)(1-\ds^2(a)) \sum_{j=1}^{p} (-1)^{j-1} d^2_j~,
\eeq
\beq\label{e80}
\sum_{j=1}^{p} (-1)^{j-1} d^3_j [d_{j+r}-d_{j-r}] =
-2m\cs(a) \sum_{j=1}^{p} (-1)^{j-1} d_j s_j c_j
+4\cs^3(a) \sum_{j=1}^{p} (-1)^{j-1} Z_j~, 
\eeq
\beq\label{e81}
\m^2\sum_{j=1}^{p} (-1)^{j-1} s^3_j [s_{j+r}-s_{j-r}] =
-2m\ns(a) \sum_{j=1}^{p} (-1)^{j-1} d_j s_j c_j
+4\ns^3(a) \sum_{j=1}^{p} (-1)^{j-1} Z_j~, 
\eeq
\beq\label{e82}
\m^2\sum_{j=1}^{p} (-1)^{j-1} c^3_j [c_{j+r}-c_{j-r}] =
-2m\ds(a) \sum_{j=1}^{p} (-1)^{j-1} d_j s_j c_j
+4\ds^3(a) \sum_{j=1}^{p} (-1)^{j-1} Z_j~, 
\eeq
\beqa\label{e83}
&&\sum_{j=1}^{p} (-1)^{j-1} d^3_j [d^3_{j+r}-d^3_{j-r}] =
4m\cs^3(a) \sum_{j=1}^{p} (-1)^{j-1} d_j s_j c_j \nonumber \\
&&-4\cs(a) \big [\ns^2(a)(\cs^2(a)+3\ds^2(a))
+\cs^2(a)(\cs^2(a)+\ds^2(a)) \big ] 
\sum_{j=1}^{p} (-1)^{j-1} Z_j~, 
\eeqa
\beqa\label{e84}
&&\m^3\sum_{j=1}^{p} (-1)^{j-1} s^3_j [s^3_{j+r}-s^3_{j-r}] =
-4m\ns^3(a) \sum_{j=1}^{p} (-1)^{j-1} d_j s_j c_j \nonumber \\
&&+4\ns(a) \big [\ns^2(a)(\cs^2(a)+\ds^2(a)+\ns^2(a))+3\cs^2(a)\ds^2(a) \big ] 
\sum_{j=1}^{p} (-1)^{j-1} Z_j~, 
\eeqa
\beqa\label{e85}
&&\m^3\sum_{j=1}^{p} (-1)^{j-1} c^3_j [c^3_{j+r}-c^3_{j-r}] =
4m\ds^3(a) \sum_{j=1}^{p} (-1)^{j-1} d_j s_j c_j \nonumber \\
&&-4\ds(a) \big [\ns^2(a)(3\cs^2(a)+\ds^2(a))+\cs^2(a)
(\cs^2(a)+\ds^2(a)) \big ] 
\sum_{j=1}^{p} (-1)^{j-1} Z_j~, 
\eeqa
\beqa\label{e86}
&&\m^2\sum_{j=1}^{p} (-1)^{j-1} d_j s_j c_j [d_{j+r} s_{j+r} c_{j+r}
-d_{j-r} s_{j-r} c_{j-r}] =
-4m\ds(a)\cs(a)\ns(a) \sum_{j=1}^{p} (-1)^{j-1} d_j s_j c_j \nonumber \\
&&+8\ds(a)\cs(a)\ns(a) [\ns^2(a)+\cs^2(a)+\ds^2(a)]
\sum_{j=1}^{p} (-1)^{j-1} Z_j~, 
\eeqa
\beq\label{e87}
\m^2\sum_{j=1}^{p} (-1)^{j-1} d_j s^2_j c_j [s_{j+r}-s_{j-r}] =
-2\ns(a) \sum_{j=1}^{p} (-1)^{j-1} d^4_j
+2\ns(a) [\ds^2(a)+1] 
\sum_{j=1}^{p} (-1)^{j-1} d^2_j~, 
\eeq

\sss
\noindent{\bf\Large Appendix G: Some Definite Integrals}
\sss

\noindent Here $K,E$ correspond to the complete elliptic integrals 
of the first and second kind
respectively.

{\small
\beq\label{g1}
\int_{0}^{2K}\!\!\!\!  \dn^3(x)\dn(x+a)\,dx 
=2\ds(a)\ns(a) E-2K \cs^2(a) [\dn(a)-\cs(a)\Z(a)],
\eeq
\beq\label{g2}
\int_{0}^{2K} \!\!\!\! \m^2\sn^3(x)\sn(x+a)\,dx 
=2\cs(a)\ds(a) E-2K[\cs(a)\ds(a)-\ns^3(a)\Z(a)],
\eeq
\beq\label{g3}
\int_{0}^{2K}\!\!\!\!  \m^2\cn^3(x)\cn(x+a)\,dx
=2\cs(a)\ns(a) E+2K[\m^2\cn(a)-\cs(a)\ns(a)+\ds^3(a)\Z(a)],
\eeq
\beq\label{g4}
\int_{0}^{2K}\!\!\!\!\!\! \m \dn(x)\sn(x)
\dn(x+a)\sn(x+a) \,dx
=4\cs(a)\ns(a) E-2K \ns(a)[\cs(a)(1+\dn^2(a))-(1+\cn^2(a))\ds(a)\ns(a)\Z(a)],
\eeq
\beq\label{g5}
\int_{0}^{2K}\!\!\!\!\!\! \m \dn(x)\cn(x)
\dn(x+a)\cn(x+a)\,dx
=-4\cs(a)\ds(a) E
+2K [2\cs(a)\ds(a)-(\cs^2(a)+\ds^2(a))\ns(a)\Z(a)],
\eeq
\beq\label{g6}
\int_{0}^{2K}\!\!\!\!\!\! \m^2 \sn(x)\cn(x)\sn(x+a)\cn(x+a)\,dx
=4\ds(a)\ns(a) E+2K \ns(a)(1+\dn^2(a))[\cs(a)\ns(a)\Z(a)-\ds(a)],
\eeq
\beqa\label{g7}
&&\f{1}{2K} \int_{0}^{2K}\!\!\!\! \dn(x)\dn(x+a)\dn(x+a')\dn(x+a'')\,dx
=\dn(a)\dn(a')\dn(a'')+\cs(a)\cs(a'-a)\cs(a''-a)\Z(a) \nonumber \\
&&~~~~~~~~~~~~~~~~~~~-\cs(a')\cs(a'-a)\cs(a''-a')\Z(a')
+\cs(a'')\cs(a''-a)\cs(a''-a')\Z(a''),
\eeqa
\beqa\label{g8}
&&\f{1}{2K} \int_{0}^{2K}\!\!\!\! \m^2 \sn(x)\sn(x+a)\sn(x+a')\sn(x+a'')\,dx
=\ns(a)\ns(a'-a)\ns(a''-a)\Z(a) \nonumber \\
&&~~~~~~~~~~~~~~~~~~~-\ns(a')\ns(a'-a)\ns(a''-a')\Z(a')
+\ns(a'')\ns(a''-a)\ns(a''-a')\Z(a''),
\eeqa
\beqa\label{g9}
&&\f{1}{2K} \int_{0}^{2K}\!\!\!\! \m^2 \cn(x)\cn(x+a)\cn(x+a')\cn(x+a'') \,dx
=\m^2 \cn(a)\cn(a')\cn(a'')+\ds(a)\ds(a'-a)\ds(a''-a)\Z(a) \nonumber \\
&&~~~~~~~~~~~~~~~~~~~-\ds(a')\ds(a'-a)\ds(a''-a')\Z(a')
+\ds(a'')\ds(a''-a)\ds(a''-a')\Z(a''),
\eeqa
\beqa\label{g10}
&&\f{1}{2K} \int_{0}^{2K}\!\!\!\! \m^2 \cn(x)\sn(x+a)\cn(x+a')
\sn(x+a'')\,dx 
=\m^2 \sn(a)\cn(a')\sn(a'')-\ds(a)\ds(a'-a)\ns(a''-a)\Z(a) \nonumber \\
&&~~~~~~~~~~~~~~~~~~~+\ds(a')\ns(a'-a)\ns(a''-a')\Z(a')
-\ds(a'')\ns(a''-a)\ds(a''-a')\Z(a''),
\eeqa
\beqa\label{g11}
&&\f{1}{2K} \int_{0}^{2K}\!\!\!\! \m \sn(x)\dn(x+a)\sn(x+a')
\dn(x+a'') \,dx
=-\ns(a)\ns(a'-a)\cs(a''-a)\Z(a) \nonumber \\
&&~~~~~~~~~~~~~~~~~~~+\ns(a')\cs(a'-a)\cs(a''-a')\Z(a')
-\ns(a'')\cs(a''-a)\ns(a''-a')\Z(a''),
\eeqa
\beqa\label{g12}
&&\f{1}{2K} \int_{0}^{2K}\!\!\!\! \m \cn(x)\dn(x+a)\cn(x+a')\dn(x+a'')\,dx 
=\m \dn(a)\cn(a')\dn(a'')+\ds(a)\ds(a'-a)\cs(a''-a)\Z(a) \nonumber \\
&&~~~~~~~~~~~~~~~~~~~-\ds(a')\cs(a'-a)\cs(a''-a')\Z(a')
+\ds(a'')\cs(a''-a)\ds(a''-a')\Z(a'').
\eeqa
}

\sss

\noindent{\bf\Large Appendix H: Some Indefinite Integrals}
\sss

\noindent We give below recursion relations expressing certain arbitrary order integrals
in terms of lower order integrals, well known integrals of 
$\sn^n (x),~\dn^n (x),~ \cn^n(x)$ \cite{gr}, and
incomplete elliptic integrals of the first, second and third
 kind, which essentially occur due to the integral
(\ref{5.17}). It should be noted that in view of the identities (\ref{a2}) and
(\ref{a3}), the integrals for $\m\sn(x+a)\sn(x)$ and $\m\cn(x+a)\cn(x)$ are
related to the integral (\ref{5.17}). In this appendix, $n \ge 1$.

{\small
\beq\label{h1}
I_{n}=\ns^2(a)I_{n-1}-\f{\m^{n-1}\ns(a)}{(2n-1)} \sn^{2n-1}(x)
- \cs(a) \ds(a) \m^{n-1} \int \, \sn^{2n-1}(x) \,dx,
\eeq
where, $I_k \equiv \int \, \m^{k} \sn^{2k}(x)\sn(x+a)\,dx.$
\sss
\beq\label{h2}
I_{n}=-\ds^2(a)I_{n-1}+\f{\m^{n-1}\ds(a)}{(2n-1)} \cn^{2n-1}(x)
+ \cs(a) \ns(a) \m^{n-1} \int \, \cn^{2n-1}(x)\,dx,
\eeq
where, $I_k \equiv \int \, \m^{k} \cn^{2k}(x)\cn(x+a) \, dx.$
\sss
\beq\label{h3}
I_{n}=\ns^2(a)I_{n-1}-\f{\m^{n-1}\ds(a)\ns(a)}{(2n-1)} \sn^{2n-1}(x)
+ \cs(a) \m^{n} \int \, \sn^{2n-1}(x) \cn^2 (x)\,dx,
\eeq
where, $I_k \equiv \int \, \m^{k} \sn^{2k}(x) \cn(x) \dn(x+a) \, dx,
~I_0  \equiv \ds(a)\int \sn(x+a) \, dx -\cs(a) \int \sn(x) \,dx.$
\sss
\beq\label{h4}
I_{n}= -\ds^2(a)I_{n-1}-\f{\m^{n-1}\ds(a)\ns(a)}{(2n-1)} \cn^{2n-1}(x)
- \cs(a) \m^{n} \int \, \cn^{2n-1}(x) \sn^2 (x) \,dx,
\eeq
where, $I_k \equiv \int \, \m^{k} \cn^{2k}(x) \sn(x)\dn(x+a) \, dx,
~~I_0  \equiv -\ns(a)\int \cn(x+a) \, dx +\cs(a) \int \cn(x) \, dx.$
\sss
\beq\label{h5}
I_{n}=-\cs^2(a)I_{n-1}-\f{\cs(a)\ds(a)}{(2n-1)} \dn^{2n-1}(x)
+ \m \ns(a)  \int \, \dn^{2n-1}(x) \cn^2 (x) \,dx,
\eeq
where, $I_k \equiv \int \, \m \dn^{2k}(x) \cn(x) \sn(x+a)\,dx~,
~I_0  \equiv -\ds(a)\int \dn(x+a) \, dx +\ns(a) \int \dn(x) \, dx.$
\sss
\beqa\label{h6}
&&I_{n}=BI_{n-1}+\f{\ds(a)\ns(a)}{(2n-1)} \dn^{2n-1}(x)
+2\ds(a)\ns(a)\sum_{k=1}^{n-1} B^{k} \f{[\dn(x)]^{2(n-k)-1}}{2(n-k)-1} 
-2\cs(a)\ds(a)\ns(a)\!\!\int \!\!\dn(x+a) \, dx \nonumber \\
&&~~~~~~~~~~~~~~+ \cs(a) [\m+2\ds^2(a)] \int \, \dn^{2n-1}(x) \, dx
-2\cs(a)\ds^2(a)\ns^2(a) \sum_{k=1}^{n-1} B^{k-1} 
\int [\dn(x)]^{2(n-k)-1} \, dx,
\eeqa
where, $I_k \equiv \int \, \m \dn^{2k}(x) \cn(x+a) \sn(x+a) \, dx,
~~B \equiv -\cs^2(a)~,
~~I_0 = -\dn(x+a).$
\sss
\beq\label{h7}
I_{n}= -\cs^2(a)I_{n-1}+\f{\cs(a)}{(2n)} \dn^{2n}(x)
+\ds(a) \ns(a) \int \, \dn^{2n}(x) \, dx~,
\eeq
where, $I_k \equiv \int \, \m \dn^{2k+1}(x) \dn(x+a) \, dx~,
~~I_0  \equiv \int \dn(x+a) \dn(x) \, dx~.$
\sss
\beq\label{h8}
I_{n}=\ns^2(a)I_{n-1}-\f{\m^{n-1}\ns(a)}{(2n)} \sn^{2n}(x)
- \m^{n-1} \cs(a)\ds(a)  \int \, \sn^{2n}(x) \, dx~,
\eeq
where, $I_k \equiv \int \m^{k+1} \sn^{2k+1}(x) \sn(x+a) \, dx~,
~I_0  \equiv \int \m \sn(x+a) \sn(x) \, dx~=
+\dc(a) x -\nc(a) \int \dn(x+a) \dn(x) \, dx.$
\sss
\beq\label{h9}
I_{n}=-\ds^2(a)I_{n-1}+\f{\m^{n-1}\ds(a)}{(2n)} \cn^{2n}(x)
+ \m^{n-1} \cs(a)\ns(a)  \int \, \cn^{2n}(x) \, dx~,
\eeq
where, $I_k \equiv \int \!\!\m^{k+1} \cn^{2k+1}(x) \cn(x+a) \, dx~,
~I_0  \equiv \int \!\! \m \cn(x+a) \cn(x) \, dx~
=-(1-\m)\nc(a) x +\ds(a) \int \!\!\dn(x+a) \dn(x) \, dx.$
\sss
\beqa\label{h10}
I_{n}=&&\!\!\!\!\!\!\!\!BI_{n-1}+\f{\ds(a) \ns(a)}{(2n)} \dn^{2n}(x) 
+\ds(a)\ns(a) \sum_{k=1}^{n-1} B^k \f{[\dn(x)]^{2(n-k)}}{(n-k)}-2\cs(a)\ds^2(a)\ns^2(a) \sum_{k=1}^{n-1} B^{k-1} 
\int [\dn(x)]^{2(n-k)} \, dx \nonumber \\
&&+ \cs(a)[\m+2\ds^2(a)]  \int \, \dn^{2n}(x) \, dx 
-2\cs(a)\ds(a)\ns(a)B^{n-1} \int \dn(x+a) \dn(x) \, dx~,
\eeqa
where, $~I_k \equiv \!\int  \m \dn^{2k+1}(x) \sn(x+a) \cn(x+a) \, dx~,~B \equiv -\cs^2(a)~,
~$
$$I_0  \equiv \!\!\int \!\!\m \dn(x) \sn(x+a) \cn(x+a) dx= -(1-\m)\cs(a) x +\cs(a) E(\am~ x,k)
+\ds(a)\nc(a) \int \dn(x+a) \dn(x) dx.
$$
}

\end{document}